\documentclass[floats,aps,prfluids,superscriptaddress,floatfix,onecolumn,nofootinbib]{revtex4-2}
\usepackage{natbib} 
\usepackage{times} 
\usepackage{amssymb,amsmath}
\usepackage{csquotes}
\usepackage{array} %
\usepackage{multirow}%
\usepackage{prelim2e}\usepackage[none,bottom]{draftcopy}
\draftcopyName{preprint / }{1.2} %
\usepackage[usenames,dvipsnames]{color}
\usepackage[pdftex]{graphicx}
\usepackage{epstopdf}
\usepackage[pdftex]{hyperref}
\hypersetup{pdftitle={Mesoscopic hydrodynamic model for spreading, sliding and coarsening compound drops} %
    pdfauthor={Jan Diekmann and Uwe Thiele},  %
pdfproducer={lateX},
pdfview=FitV,       %
pdfstartview=FitB,
linkcolor=blue,     %
citecolor=blue,     %
urlcolor=red,      %
breaklinks=true,    %
colorlinks=true,
citebordercolor=0 0 0,  %
filebordercolor=0 0 0,
linkbordercolor=0 0 0,
menubordercolor=0 0 0,
urlbordercolor=0 0 0,
pdfhighlight=/I,
pdfborder=0 0 0,   %
bookmarksopen=true,
bookmarksnumbered=true
}
\DeclareGraphicsExtensions{.jpg, .pdf, .tif, .png}
\graphicspath{{.}}%
\usepackage[normalem]{ulem}
\usepackage{tcolorbox}

\renewcommand{\vec}[1]{\mathbf{#1}}

\newcommand{\tens}[1]{\mathbf{\underline{#1}}}

\usepackage{wrapfig}
\usepackage{mathtools}
\usepackage[section]{placeins}
\usepackage{xcolor}
\newcommand{\xiontw}{\xi_\text{12}}
\newcommand{\xitwth}{\xi_\text{23}}
\newcommand{\xionth}{\xi_\text{13}}
\newcommand{\honth}{h_\text{13}}
\newcommand{\htwth}{h_\text{23}}
\newcommand{\hontw}{h_\text{12}}
\newcommand{\lone}{L_1}
\newcommand{\ltwo}{L_2}
\newcommand{\young}[1]{\theta_\mathrm{#1}}
\newcommand{\gam}[1]{\gamma_\mathrm{#1}}
\newcommand{\neum}[1]{\psi_{#1}}
\newcommand{\gsth}{\gam{s3}}
\newcommand{\gstw}{\gam{s2}}
\newcommand{\gson}{\gam{s1}}
\newcommand{\gontw}{\gam{12}}
\newcommand{\gonth}{\gam{13}}
\newcommand{\gtwth}{\gam{23}}
\newcommand{\gsg}{\gamma_\text{s3}}
\newcommand{\gsl}{\gamma_\text{s1}}

\newcommand*{\haA}[0]{h_{\mathrm{a}1}}
\newcommand*{\haB}[0]{h_{\mathrm{a}2}}
\newcommand*{\hhaB}[0]{\hat{h}_{\mathrm{a}2}}

\begin{document}
\title{Mesoscopic hydrodynamic model for spreading, sliding and coarsening compound drops}
\author{Jan Diekmann}
\email{jan.diekmann@uni-muenster.de}
\thanks{ORCID ID: 0009-0005-4741-3368}
\affiliation{Institute of Theoretical Physics, University of M\"unster, Wilhelm-Klemm-Str.\ 9, 48149 M\"unster, Germany}
\author{Uwe Thiele}
\email{u.thiele@uni-muenster.de}
\homepage{http://www.uwethiele.de}
\thanks{ORCID ID: 0000-0001-7989-9271}
\affiliation{Institute of Theoretical Physics, University of M\"unster, Wilhelm-Klemm-Str.\ 9, 48149 M\"unster, Germany}
\affiliation{Center for Nonlinear Science (CeNoS), University of M\"unster, Corrensstr.\ 2, 48149 M\"unster, Germany}
\affiliation{Center for Multiscale Theory and Computation (CMTC), University of M\"unster, Corrensstr.\ 40, 48149 M\"unster, Germany}
\begin{abstract} We revisit the mesoscopic hydrodynamic description of the dynamics of sessile {partially wetting} compound drops, i.e., of drops that consist of two immiscible nonvolatile partially wetting liquids and are situated on a smooth rigid solid substrate. We briefly discuss and complete existing {dynamic} models employing a gradient dynamics approach. {Thereby, the underlying energy features capillarity and wettability contributions for all relevant interfaces in full-curvature formulation. Establishing transparent consistency relations between macroscopic and mesoscopic parameters, we obtain} mesoscopic Neumann and Young laws that are also fully consistent with the macroscopic ones. In particular, we discuss the minimal requirements for the wetting energy that ensure the full spectrum of macroscopic parameters {for partially wetting cases} is addressed by the mesoscopic model. {Subsequently, we distinguish long-wave and full-curvature variants of the dynamical model based on properties of the energy, and employ the latter} to illustrate the usage of the mesoscopic model. As examples, we {chose} the spreading of individual compound drops on one-dimensional horizontal substrates, sliding compound drops on one-dimensional inclined substrates, and the coarsening of drop ensembles on one- and two-dimensional horizontal substrates. In each case, the discussion emphasizes occurring qualitative changes in the drop configurations.
\end{abstract}
\maketitle
\begin{tcolorbox}[arc=0mm,colback=black!5!white,colframe=black!30!red,title=Publication note]
This version of the article has been accepted for publication, after peer review but
is not the Version of Record and does not reflect all post-acceptance improvements and
corrections. The Version of Record is available online at\newline
\begin{center}
Diekmann and Thiele, Physical Review Fluids 10, 024002, 2025, doi:  \href{https://doi.org/10.1103/PhysRevFluids.10.024002}{10.1103/PhysRevFluids.10.024002}.
\end{center}%
\end{tcolorbox}
\section{Introduction} \label{sec:intro}

Bilayer liquid films and compound drops of two immiscible liquids on solid substrates have over time attracted considerable interest. Studied phenomena include, for instance, the dewetting of two-layer films of immiscible liquids \cite{BrMR1993l,FaCW1995l,SHCB1998prl,PBMT2004pre,PBMT2005jcp,BaGS2005iecr,BaSh2006jcp,YaKK2012prb,BCJP2013epje,PBJS2018sr}, the spreading of drops on another liquid \cite{Lued1869app,CrMa2006jcis}, the development of long-wave thermocapillary or electro-static instabilities in such bilayer films \cite{Badr1985igsansn,BaSh2007jcis,NeSi2007pf,NeSi2009prl,NeSi2017pf,NeSi2021prf}, the occurrence of different steady configurations of sessile compound drops of two (or more) liquids \cite{MaAP2002jfm,NTGD2012sm,BSNB2014l,ZCAG2016jcis,IDSS2017l,Kita2024jons}, and the capillary leveling of two-layer films \cite{BLSR2021jfm}. More involved systems are also investigated as, for instance, heated bilayer systems with soluble or insoluble surfactants and evaporation \cite{DPAR1998ces,DPSA1998ces,PDAR1998ces}, sliding drops on liquid layers \cite{KrMi2003sjam}, and bilayer films of viscoelastic and elastic materials \cite{MuSh2015sm}. The decomposition of a film of a liquid mixture into a two-layer film that may at the same time dewet is also considered \cite{GeKr2003pps,GRBP2006pla}. The influence of slip is discussed in \cite{JPMW2014jem,XBRS2017sr}. Finally note that the description of drops on liquid-infused substrates is also closely related to two-layer systems \cite{GWXM2015l,SeMK2017sm,TKPS2017sm}.

Although descriptions of the dynamics of two-layer films with the full Navier-Stokes equations are possible \cite{VNDS2010jfm}, most modeling efforts focus on thin-film (or lubrication, or long-wave) models \cite{OrDB1997rmp,CrMa2009rmp} that allow one to consider a reduced dynamics in the form of coupled time evolution equations for {either the thickness profiles of the two liquid layers or the height profiles of the two free interfaces} \cite{Kita2024jons,BrMR1993l,PBMT2004pre,BaGS2005iecr,CrMa2006jcis,BaSh2006jcp,Ward2011pf,BCJP2013epje,JHKP2013sjam}.

Here, we focus on the isothermal case {of} thin films {and droplets} of maximal thicknesses well below the capillary length, i.e., we neglect hydrostatic influences and fully concentrate on a mesoscopic hydrodynamic description that faithfully incorporates substrate wettability and fluid-fluid interface energies. Such a description is commonly employed to study, for instance, the rupture of bilayer films \cite{PBMT2004pre,Ward2011pf}, the initial dewetting together with the later drop coarsening and related morphology changes for one-dimensional \cite{PBMT2005jcp} and two-dimensional \cite{PBMT2006el} substrates, the dewetting of a liquid layer and the subsequent dynamics of liquid droplets on a stable layer of another liquid \cite{BCJP2013epje}, and the behavior of liquid lenses \cite{CrMa2006jcis}. As the underlying long-wave approximation assumes a small ratio of typical length scales orthogonal and parallel to the solid substrate, the resulting models are normally limited to small slopes of all involved interfaces. However, below we will also discuss a full-curvature variant of such models {that combine long-wave dynamics with full-curvature energetics. For one-layer systems this approach is discussed, e.g.,} in Refs.~\cite{Snoe2006pf,BoTH2018jfm,Thie2018csa} {as well as in appendix~A of \cite{ThAP2016prf}}. This is closely related to the possibility {of formulating all mentioned mesoscopic models \cite{BrMR1993l,PBMT2004pre,BaGS2005iecr,CrMa2006jcis,BaSh2006jcp,Ward2011pf,BCJP2013epje,JHKP2013sjam} as gradient dynamics on an underlying energy functional \cite{PBMT2004pre,PBMT2005jcp,PBMT2006el,BCJP2013epje,HJKP2015jem,PBJS2018sr,Kita2024jons}. Note, however, that some of these formulations need to be carefully scrutinized to uncover their gradient dynamics form.}

For static situations, e.g., sessile compound drops, a macroscopic description is also commonly used \cite{MaAP2002jfm,NTGD2012sm,ZCAG2016jcis,IDSS2017l}. Then, the system can be completely characterized by the involved fluid-solid and fluid-fluid interface energies. All involved fluid-fluid interfaces are of constant curvature governed by respective Laplace laws. All three-phase contact lines where three fluid-fluid interfaces meet correspond to circular arcs, and the angles between the interfaces are governed by Neumann's law (chap.~6.1 and 6.2 of \cite{NeumannWangerin1894}, also compare \cite{BoisReymond1859}). For three fluids including the ambient gas or liquid there are three fluid-fluid and three fluid-solid interface energies. That is, macroscopically there are six energetic parameters of which two can be eliminated by general arguments (see below). For a complete nondimensional macroscopic description, one has to add one volume ratio.\footnote{Note that here we do not take into account that the macroscopic interface energies might be related among themselves by microscopic considerations, e.g., in the case of purely apolar fluids by sum rules as, e.g., discussed in chapter~10 of \cite{Israelachvili2011}.}

Note that other approaches exist that, e.g., employ a phase-field model \cite{BSNB2014l} or use piecewise thin-film descriptions for the individual interfaces coupled at sharp contact lines \cite{KrMi2003sjam,PBJS2018sr}. However, then the equilibrium Neumann law is imposed at the triple line \cite{KrMi2003sjam} what might be problematic in dynamic situations. A corresponding comparison of a mesoscopic thin-film model (that allows for adsorption layers, i.e., sharp three-phase contact lines are replaced by smooth transition regions) and a piecewise model for relatively slow relaxational dynamics is provided by \cite{HJKP2015jem}. The deviation from Neumann's law in the dynamic situation of forced wetting is discussed in \cite{GrHT2023sm} for a polymer brush-covered substrate that is moved at given velocity into a liquid bath.

The aim of the present work is to present a mesoscopic hydrodynamic model for {compound drops of two immiscible nonvolatile partially wetting liquids on smooth rigid solid substrates that is based on a gradient dynamics approach. It shall be} sufficiently general to (i) describe {the initial two-layer dewetting dynamics} and the subsequent coarsening process, (ii) capture all types of steady compound drops described in the literature as well as their dynamics (e.g., spreading, sliding, coarsening), and (iii) allow for consistency with all relevant macroscopic Laplace, Neumann and Young laws {in the case of partial wetting}. Furthermore, {a long-wave variant and a full-curvature variant will be discussed based on different treatments of the underlying energy functional.}

We argue that such a re-visitation is necessary as the mesoscopic models in the literature are normally not of a form that can be easily related to the appropriate macroscopic description, and do not allow for a consistent analysis of sessile compound drops in the full range of macroscopic parameters. For instance, although the two-layer models of Ref.~\cite{PBMT2005jcp,PBMT2006el} are able to capture a number of compound drops (see e.g., Fig.~14 of \cite{PBMT2005jcp}), the employed form of the wetting and interface energy does not lend itself easily to a clear discussion of consistency conditions between macroscopic and mesoscopic modeling or, indeed, to the discussion of a mesoscopic equivalent of Neumann's law. The two-layer models that focus on the initial stages of the dewetting and rupture process are not able to describe sessile compound drops because no stabilizing short-range interactions are included in the wetting energy \cite{PBMT2004pre,Ward2011pf}. Other works assume the liquid of the lower layer to completely wet the substrate, i.e., again, no compound drops can emerge \cite{CrMa2006jcis,BCJP2013epje,PBJS2018sr}. These restrictions also apply to other more complicated models that involve more components beside the two liquids, e.g., surfactants, further liquids, and/or evaporation \cite{DPAR1998ces}.

The mesoscopic dynamic model is provided in section~\ref{sec:model:dynamics} in the form of a gradient dynamics on a mesoscopic energy functional that incorporates wetting and interface energies in the full-curvature variant. {A corresponding long-wave variant} is also given. The subsequent section~\ref{sec:statics} discusses important aspects related to the form of the energy functional, and shows that the resulting mesoscopic Laplace, Neumann and Young laws all converge to the correct macroscopic expressions if certain consistency conditions are fulfilled. Details of the related energy minimizations and resulting comparisons of mesoscopic and macroscopic Laplace, Young and Neumann laws are given in appendix~\ref{app:neumann}. The full-curvature variant of the presented model is then employed for a few examples in section~\ref{sec:appl}: First, we investigate in section~\ref{sec:spreading} the spreading of compound drops, then in section~\ref{sec:sliding} we consider how the configuration of sliding compound drops may change, and in section~\ref{sec:coarsening} we illustrate the dewetting and coarsening behavior of a two-layer film. Finally, section~\ref{sec:conc}  concludes and gives an outlook. 

\section{Gradient dynamics model}
\label{sec:model:dynamics}
\begin{figure}[tbh]
	\includegraphics[width=0.9\textwidth]{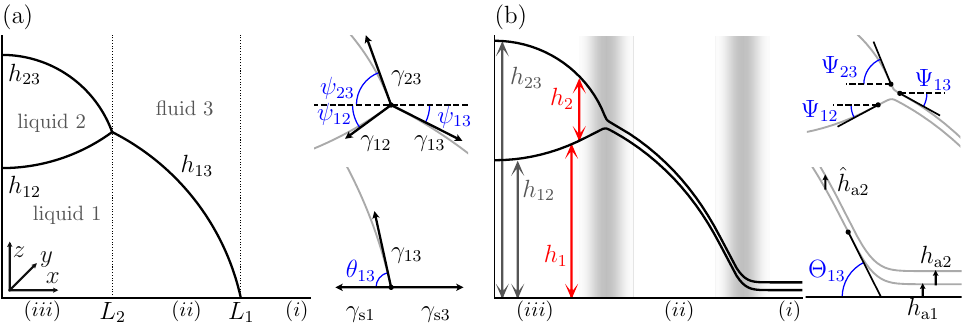}\\\vspace*{10pt}
	\includegraphics[width=0.9\textwidth]{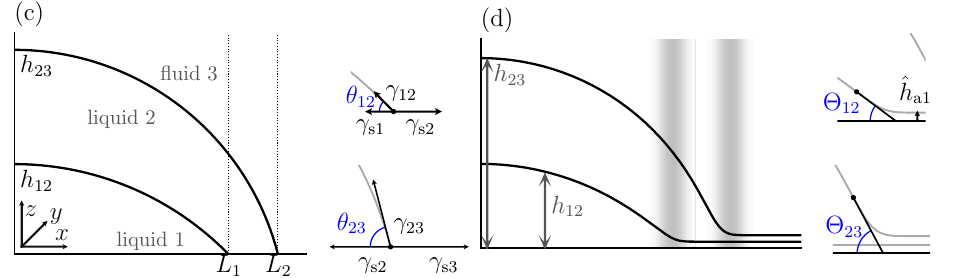}
	\caption{Sketches illustrate {(a,c) macroscopic compound drops and (b,d) their mesoscopic equivalents where thin adsorption layers are present. Shown are (a,b) a drop-on-drop and (c,d) a drop-covers-drop configuration where adsorption layers occur at (b) the liquid~1-gas and the solid-gas interface, and (d) at the solid-liquid~2 and solid-gas interface. Macroscopic three-phase contact-points in (a) and (c) translate into mesoscopic three-phase contact regions in (b) and (d), respectively (indicated by gray shading).} Magnifications of the relevant regions are given and illustrate corresponding Neumann and Young constructions in their macroscopic {[(a) and (c)]} and mesoscopic {[(b) and (d)] form.} Here, mesoscopic contact angles are determined at inflection points of the respective profiles. Note that the mesoscopic profiles in {(b) and (d)} represent actual numerically obtained steady states and are connected to the macroscopic profiles in {(a) and (c)} by the consistency relations discussed in the main text. Furthermore, panel~(b) indicates the two different choices for the considered profiles: either in terms of layer thickness profiles $h_1$ and $h_2$ (red arrows) or in terms of interface height profiles $h_{12}$ and $h_{23}$ (gray arrows). 
	}
	\label{fig:sketch_macro_meso}
\end{figure}
We employ the governing equations for the time evolution of the bottom and top layer thickness profiles $h_1(\vec{r},t)$ and $h_2(\vec{r},t)$, respectively, or alternatively, for the interface height profiles $h_{12}(\vec{r},t)$ and $h_{23}(\vec{r},t)$, see {sketches in} Fig.~\ref{fig:sketch_macro_meso}. They are written in standard gradient dynamics formulation for two scalar fields with conserved dynamics \cite{Kita2024jons,PBMT2004pre,PBMT2005jcp,PBMT2006el,BCJP2013epje,JHKP2013sjam,HJKP2015jem,PBJS2018sr,Thie2018csa}, e.g., for the layer thicknesses
	\begin{align}\label{eq:pde_system}
		\begin{split}
		\partial_t h_1 &=- \nabla\cdot\vec{j}_1=\nabla\cdot \left(Q_{11}\nabla\frac{\delta \mathcal{F}}{\delta h_1} + Q_{12}\nabla\frac{\delta \mathcal{F}}{\delta h_2}\right)\\
		\partial_t h_2 &=- \nabla\cdot\vec{j}_2=\nabla\cdot \left(Q_{21}\nabla\frac{\delta \mathcal{F}}{\delta h_1} + Q_{22}\nabla\frac{\delta \mathcal{F}}{\delta h_2}\right)\,, 
		\end{split}
	\end{align}
        where $\mathcal{F}$ is the underlying energy functional and $\tens{Q}$ is a symmetric, positive definite mobility matrix ensuring positive entropy production and Onsager's reciprocity relations. The positions on the two-dimensional planar solid rigid substrate are indicated by $\vec{r}=(x,y)^T$, time is $t$, $\partial_t$ stands for the partial time derivative, and $\nabla=(\partial_x, \partial_y)^T$.  Furthermore, $\vec{j}_1$ and $\vec{j}_2$ are the lateral fluxes within the respective liquid layer.  For the presently used variables ($h_1$ and $h_2$) one has \cite{BCJP2013epje,JPMW2014jem}
\begin{align}\label{eq:mobilitymatrix}
	\tens{Q} = 
	\begin{pmatrix}
		Q_{11} & Q_{12}\\
		Q_{21} & Q_{22}
	\end{pmatrix}
	=\frac{1}{6\eta_1}\begin{pmatrix}
	2h_1^3& 3h_1^2 h_2 \\
	3h_1^2 h_2 &2\frac{\eta_1}{\eta_2} h_2^3 + 6 h_1 h_2^2
	\end{pmatrix},
\end{align}
where $\eta_1$ and $\eta_2$ are the dynamic viscosities of bottom and top liquid, respectively. {These mobilities are} determined from a long-wave (lubrication) approximation of the governing hydrodynamic equations and appropriate boundary conditions (no-slip and no-penetration at solid substrate, velocity and shear stress continuity at liquid-liquid interface, zero shear stress at liquid-gas interface) \cite{PBMT2004pre,PBMT2005jcp,JHKP2013sjam} following the general procedure as outlined in the reviews~\cite{OrDB1997rmp,Thie2007chapter,CrMa2009rmp}. For a case with slip at the solid substrate see Ref.~\cite{PBMT2005jcp}. Slip at liquid-solid and liquid-liquid interface is incorporated in \cite{JPMW2014jem}.

The employed energy functional is 
\begin{equation}
  \label{eq:energy-meso}
\mathcal{F} \,=\, \int_A \left[ \gamma_{\mathrm{s}1}+\gamma_{12} \,\xi_{12} + \gamma_{23}  \,\xi_{23} + g(h_1,h_2,\xi_{23}) \right]\, \mathrm{d}^2r\,.
\end{equation}
{In the full-curvature variant we use the exact} metric factors $\xi_{12}=\sqrt{1+|\nabla h_1|^2}$ and $\xi_{23}=\sqrt{1+|\nabla (h_1+h_2)|^2}$, {while they are approximated in the long-wave variant.\footnote{{The sometimes conceived conflict between mathematically proper asymptotic usage of smallness parameters as in the long-wave variant (that then in practice are often not small) and an approximation scheme where the thermodynamic structure of the model is employed (and always preserved as the \enquote{higher good}) in approximation schemes like the full-curvature variant is discussed, e.g., in Refs.~\cite{Snoe2006pf,BoTH2018jfm,Thie2018csa} and appendix~A of \cite{ThAP2016prf}. We advocate the latter and independently approximate mobilities and energies (like it is common for, e.g., the Cahn-Hilliard equation where mobilities are often set constant and energies are expanded up to fourth order).}}} The first three terms correspond to the interface energies of the solid-liquid, liquid-liquid and liquid-gas interface, respectively (with energy densities $\gamma_{\mathrm{s}1}, \gamma_{12}$ and $\gamma_{23}$), while the fourth term collects all parts of the wetting energy. The particular form of $g(h_1,h_2,\xi_{23})$ including its dependence on a metric factor is discussed below in section~\ref{sec:statics}. {As in the following we consider an isothermal situation, the energy $\mathcal{F}$ may be seen as a thermodynamic free energy.}

The variations of $\mathcal{F}[h_1,h_2]$ in Eqs.~\eqref{eq:pde_system} correspond to the (vertically uniform) pressures within the respective layers
{
  \begin{equation}
\frac{\delta \mathcal{F}}{\delta h_1} = p_1(\vec{r},t)\quad\mathrm{and}\quad
\frac{\delta \mathcal{F}}{\delta h_2} = p_2(\vec{r},t)\,.
	\label{eq:BCP-ENRG}
      \end{equation}
      }
{Their gradients act as thermodynamic forces that drive the dynamics \eqref{eq:pde_system}.}
{In Eqs.}~\eqref{eq:pde_system} with \eqref{eq:mobilitymatrix} and \eqref{eq:BCP-ENRG} {we denote} the full-curvature variant of the mesoscopic hydrodynamic model for a two-layer system written in gradient dynamics form. The long-wave variant {uses the approximation of the $\xi$'s and } is discussed in section \ref{sec:long-wave-meso}. 

An advantage of the presented gradient dynamics form is the resulting straightforward way of changing the describing variables \cite{XuTQ2015jpcm,ThAP2016prf}: Instead of writing the model in layer thicknesses $h_{1}$ and $h_{2}$ \cite{BCJP2013epje,HJKP2015jem,PBJS2018sr}, alternatively, one may use the height profiles of the two free interfaces $h_{12}=h_1$ and $h_{23}=h_1 + h_2$ \cite{PBMT2004pre,PBMT2005jcp,JHKP2013sjam,JPMW2014jem}. The two representations are related by a linear transformation 
\begin{align}\label{eq:h_trafo}
	\tilde h_\alpha = \sum_\beta R_{\alpha\beta} h_\beta
\end{align}
where here $(\tilde h_1, \tilde h_2)$ stands for $(h_{12}, h_{23})$ and the $2\times2$ transformation matrix is $\tens{R}=((1,0),(1,1))$.
As the variations transform as 
\begin{align}\label{eq:dh_trafo}
	\frac{\delta}{\delta h_\alpha}= \sum_\beta R_{\beta \alpha} \frac{\delta}{\delta \tilde h_\beta}\,,
\end{align} 
one obtains from Eqs.~\eqref{eq:pde_system} the transformed set of governing equations
\begin{equation}
  \partial_t \tilde h_\alpha= -\nabla\cdot\tilde{\vec{j}}_\alpha =\nabla\cdot \sum_{\beta=1}^2\left[\widetilde Q_{\alpha\beta}(\tilde h_1,\tilde h_2) \nabla\frac{\delta      \mathcal{F}}{\delta \tilde h_\beta}\right]
  \quad\mathrm{for}\quad \alpha=1,2
\label{eq:trans-gen}
\end{equation}
where the transformed (again symmetric positive definite) mobility matrix is $\widetilde{\tens{Q}}=\tens{R}\tens{Q}\tens{R}^T$. The lateral fluxes $\tilde{\vec{j}}_\alpha$ transform as the profiles, i.e., as in Eq.~\eqref{eq:h_trafo}.
Specifically, one has \cite{PBMT2004pre,PBMT2005jcp,JHKP2013sjam,JPMW2014jem}
\begin{align}\label{eq:mobilitymatrix-alt}
	\widetilde{\tens{Q}} =\frac{1}{\eta_1}\begin{pmatrix}
		\frac{\hontw^3}{3} & \frac{\hontw^2}{2}\left(\htwth - \frac{\hontw}{3}\right)\\
		\frac{\hontw^2}{2}\left(\htwth - \frac{\hontw}{3}\right) & \frac{(\htwth-\hontw)^3}{3}\left(\frac{\eta_1}{\eta_2}-1\right) +  \frac{\htwth^3}{3}
	\end{pmatrix}.
\end{align}
Note that the transformation is also applied within the energy functional. Then, the two alternative formulations are fully equivalent. After reformulating by using Eq.~\eqref{eq:dh_trafo}, i.e., $\delta/\delta \tilde{h}_\alpha=\sum_\beta R^{-1}_{\beta \alpha} \delta/\delta {h}_\beta$, the variations are
\begin{eqnarray}
\begin{split}
\frac{\delta \mathcal{F}}{\delta h_{12}} &= \frac{\delta \mathcal{F}}{\delta h_{1}} -\frac{\delta \mathcal{F}}{\delta h_{2}} = p_1-p_2 \\
\frac{\delta \mathcal{F}}{\delta h_{23}} &= \frac{\delta \mathcal{F}}{\delta h_{2}} = p_2 - p_3
\end{split}
	\label{eq:BCP-ENRG2}
\end{eqnarray}
and correspond to pressure jumps across interfaces. 
The pressure in the gas phase $p_3$ is normally neglected in the case of nonvolatile liquids.

Note that beside the two alternative gradient dynamics formulations, in terms of layer thicknesses $h_1$ and $h_2$ (Eqs.~\eqref{eq:pde_system}) as used, e.g., in Refs.~\cite{Kita2024jons,BCJP2013epje,JPMW2014jem,HJKP2015jem,PBJS2018sr}, and in terms of interface heights $h_{12}$ and $h_{23}$ (Eq.~\eqref{eq:trans-gen}) as used, e.g., in Refs.~\cite{PBMT2004pre,PBMT2005jcp,KPMW2011cep,JHKP2013sjam,JPMW2014jem}, there exists also a mixed formulation, see Refs.~\cite{FiGo2005jcis,BaGS2005iecr,BaSh2006jcp,CrMa2006jcis,FiGo2007jcis,NeSi2007pf,NeSi2009prl,NeSi2009jfm,NeSi2021prf}. There, the time derivatives are written in terms of $h_{12}$ and $h_{23}$, but the fluxes are given in terms of the gradients of the pressures $p_1$ and $p_2$ within the layers, and not in terms of gradients of pressure jumps $p_1-p_2$ and $p_2 - p_3$ across the interfaces at $h_{12}$ and $h_{23}$, respectively. Although the resulting models are still mathematically correct, the mixed formulation should be avoided as it obfuscates the underlying gradient dynamics structure. For instance, the resulting mobility matrix is not symmetric anymore.\footnote{
	The transformation $\vec{h}=\tens{R}^{-1}\vec{\tilde{h}}$ is applied only on the left hand side of Eq.~\eqref{eq:pde_system}, therefore, $ \partial_t \vec{\tilde{h}} = \nabla\cdot\left[\widehat{\tens{Q}}\nabla\frac{\delta\mathcal{F}}{\delta \vec{{h}}}\right]$ with the asymmetric mobility matrix\\ $\widehat{\tens{Q}}=\tens{R}\tens{{Q}}=\frac{1}{6\eta_1}
		\begin{pmatrix}
					 2\tilde{h}_1^3
					 &
					 3 \tilde{h}_1^2( \tilde{h}_2 - \tilde{h}_1)
					 \\
						\tilde{h}_1^2(3\tilde{h}_2-\tilde{h}_1)
					 &
					 (\tilde{h}_2-\tilde{h}_1)[3\tilde{h}_1^2 + 6\tilde{h}_1(\tilde{h}_2-\tilde{h}_1) + 2\frac{\eta_1}{\eta_2}(\tilde{h}_2-\tilde{h}_1)^2 ]
		\end{pmatrix}$
               where again $(\tilde h_1, \tilde h_2)$ stands for $(h_{12}, h_{23})$.}
This is inconvenient as it can not be employed to easily express the linear stability analysis as a generalized eigenvalue problem (cf.~\cite{PBMT2005jcp}) rendering the analysis quite cumbersome. Furthermore, in this formulation it may appear that Derjaguin pressures do not result from a single underlying wetting potential.

\section{Steady states}
\label{sec:statics}

First, we consider static situations, i.e., steady states described by Eqs.~\eqref{eq:pde_system} with $\partial_t h_1=\partial_t h_2=0$. Integrating the resulting equations once, the integration constants are set to zero for the considered situation without fluxes across the lateral domain boundaries. Taking then into account that the mobility matrix $\tens{Q}$ is positive definite for $h_1, h_2 > 0$, i.e., $\det \tens{Q} >0$, one can integrate a second time. {One} obtains the condition for steady states, namely, that the variations of $\mathcal{F}$ given by Eq.~\eqref{eq:BCP-ENRG} are uniform in space, i.e., the pressures are equal to constants $P_1$ and $P_2$ that appear in the second integration.
Equations \eqref{eq:BCP-ENRG} then correspond to coupled nonlinear ordinary differential equations for $h_1$ and $h_2$ that may be solved, e.g., by numerical continuation as done in Ref.~\cite{PBMT2005jcp}, see their Figs.~8, 9, 12-14.  

Resulting steady compound drops may show three-phase contact regions where three media meet, see Fig.~\ref{fig:sketch_macro_meso}(b),(d). On the one hand, this may be the liquid-liquid-gas contact of three fluids where macroscopically the Neumann law holds. On the other hand, this may be a solid-liquid-gas or a solid-liquid-liquid contact where two fluids and the solid substrate meet. There, macroscopically a respective Young law holds. The derivation of these macroscopic laws via minimization of a macroscopic equivalent of the energy~\eqref{eq:energy-meso} is briefly reviewed in Appendix~\ref{sec:statics-macro}. Here, we next derive the mesoscopic version of these laws and subsequently provide consistency relations connecting the mesoscopic and macroscopic descriptions.

\subsection{Mesoscopic Laplace, Neumann and Young laws}
\label{sec:statics-laws}

In this section, we restrict the consideration to one-dimensional substrates and use the formulation in {interface} height profiles $h_{12}$ and $h_{23}$. We start with the mesoscopic energy functional 
\begin{align}
	\begin{split}
		\mathcal{F}_\text{meso} =
		\int_{-\infty}^{\infty} &\left(\gsl + \gontw\xiontw + \gtwth \xitwth + f_1(\hontw) + \xitwth f_2\left(\htwth-\hontw\right) + f_3(\htwth)\right.\\
		&\,\left.\vphantom{\frac{1}{1}} -P_1(\hontw -\bar{h}_1) - P_2(\htwth-\hontw-\bar{h}_2)\right)\, \mathrm{d}x \,,
 	\end{split}\label{eq:meso_grandpotential}
\end{align}
where we have directly included the constraints for volume conservation
{\begin{align}
	\int_{-\infty}^{\infty} (\hontw-\bar{h}_1)\,\mathrm{d}x &= 0\\ %
	\text{and } \quad \int_{-\infty}^{\infty} (\htwth-\hontw-\bar{h}_2)\mathrm{d}x &= 0 %
       \end{align}
       }
       with respective Lagrange multipliers $P_1$ and $P_2$, and respective mean layer thicknesses $\bar{h}_1$ and $\bar{h}_2$. {The energy $\mathcal{F}_\text{meso}$ may still be seen as a thermodynamic Helmholtz free energy as the constraints for volume conservation do not contribute when introducing \eqref{eq:meso_grandpotential} into the mass-conserving dynamics \eqref{eq:pde_system}. However, when considering steady states the constraints matter as they control the volume and one may see the functional as a Gibbs free energy.}

Note that in Eq.~\eqref{eq:meso_grandpotential} we have specified the general wetting energy using
\begin{equation} \label{eq:gg}
  g(h_1,h_2,\xi_{23})=f_1(\hontw) + \xitwth f_2\left(\htwth-\hontw\right) + f_3(\htwth),
\end{equation}
i.e., the sum of the interactions of liquid 2 with the substrate across liquid 1 ($f_1$), of the ambient gas with liquid~1 across liquid~2 ($f_2$), and of the ambient gas with the substrate across liquids~1 and~2 ($f_3$). {To capture partially wetting liquids we assume that each of the three $f_i$ features a single minimum at some microscopically small thickness and approaches zero for $h\to\infty$. Below, for specific calculations we further restrict this to the combination of two specific antagonistic power laws. Situations where one of the two liquids is completely wetting are not considered, but see, e.g., Refs.~\cite{BCJP2013epje,PBJS2018sr} for drops of partially wetting liquid on a liquid substrate, and Ref.~\cite{MPBT2005pf} for partially wetting sessile drops underneath a second (wetting) liquid.}

  {It is instructive to consider an analogy between steady state compound drops and Newtonian point mechanics: Interface heights of steady states that vary with position along the substrate are seen as coordinates of pseudo point particles that vary with time. Then the  integrand of the energy functional $\mathcal{F}_\text{meso}$ can be seen as a Lagrangian $\mathcal{L}$. This then allows one to define and determine the corresponding Hamiltonian $\mathcal{H}$ within the mechanical analogon. It represents a quantity (not a usual thermodynamic potential) that is uniform in space when discussing the thermodynamic steady states. In the mechanical analogon the constraint terms in Eq.~\eqref{eq:meso_grandpotential} represent external force fields.}

The corresponding Hamiltonian is given by
\begin{equation}
\mathcal{H}=\sum_j \frac{\partial \mathcal{L}}{\partial(\partial_x \tilde h_j)}(\partial_x \tilde h_j) - \mathcal{L}\,,
\end{equation}
where $\frac{\partial \mathcal{L}}{\partial(\partial_x \tilde h_j)}$ and $\tilde h_j$ are generalized momenta and generalized coordinates, respectively (with $\tilde h_1=\hontw$ and $\tilde h_2=\htwth$ as before). The Hamiltonian reads explicitly
\begin{align}
	\mathcal{H} &= 
	\gontw \dfrac{\left(\partial_x\hontw\right)^2}{\xiontw}	+ (\gtwth + f_2)  \dfrac{(\partial_x\htwth)^2}{\xitwth}	- \mathcal{L}\\
		&= -\frac{\gontw}{\xiontw}  		- \frac{\gtwth+f_2}{\xitwth}		- \gsl - f_1 -f_3 + P_1(\hontw-\bar{h}_1) + P_2(\htwth-\hontw-\bar{h}_2)\,.\label{eq:mesoI_hamiltonian}
\end{align}
Note that from hereon we use the energy $E=-\mathcal{H}$. Minimization of $\mathcal{F}_\text{meso}$ w.r.t.\ $\hontw$, and $\htwth$ gives the Euler-Lagrange equations 
\begin{align}
	 -\gontw \kappa_{12} + f_1^\prime - \xitwth f_2^\prime =& P_1-P_2\label{eq:mesoI_eula1}\,,
\end{align}
and
\begin{align}
		-\left(\gtwth 		+f_2\right)\kappa_{23}
		+ \left(\xitwth-\frac{(\partial_x\htwth-\partial_x\hontw)\partial_x\htwth}{\xitwth}\right) f_2^\prime + f_3^\prime =& P_2,
\label{eq:mesoI_eula2}
\end{align}
respectively, with curvatures $\kappa_{12} = (\partial_{xx} h_{12})/\xi^3_{12}$ and $\kappa_{23} = (\partial_{xx} h_{23})/\xi^3_{23}$. In the mechanical {analogon} they correspond to coupled Newton's equations of motion for particles with \enquote{velocity- and interaction-dependent masses} $\gontw /\xi^3_{12}$ and $(\gtwth+f_2)/\xi^3_{23}$. 

To obtain mesoscopic Laplace, Neumann and Young laws we consider three regions (i), (ii) and (iii) at large (macroscopic) distance from the two distinguished contact regions (macroscopically located at $\lone$ and $\ltwo$, cf.~Fig.~\ref{fig:sketch_macro_meso}{(a) and~\ref{fig:sketch_macro_meso}(b)}) and consider the corresponding limiting cases of Eqs.~\eqref{eq:mesoI_hamiltonian}, \eqref{eq:mesoI_eula1} and \eqref{eq:mesoI_eula2}.
First, we consider region (i) where $x\gg\lone$, i.e., $x\rightarrow\infty$. There, the substrate is only covered by ultrathin adsorption layers of uniform thicknesses $\haA$ and $\haB$, implying that Eqs.~\eqref{eq:mesoI_eula1} and \eqref{eq:mesoI_eula2} become
after rearrangement
	\begin{align}
		f_1^\prime(\haA) +f_3^\prime(\haA+\haB) & = P_1, \label{eq:meso_limit_i_condition1}\\
		f_2^\prime(\haB)+f_3^\prime(\haA+\haB) & = P_2 \,. \label{eq:meso_limit_i_condition2}
	\end{align} 
Similarly, Eq.~\eqref{eq:mesoI_hamiltonian} becomes 
\begin{align}\label{eq:meso_limit_i_condition3}
	E = 	\gsl + \gontw + \gtwth + f_1(\haA) + f_2(\haB)+ f_3(\haA+\haB)   - P_1 (\haA-\bar{h}_1) - P_2 (\haB-\bar{h}_2)\,.
\end{align}

Second, we consider the intermediate region (ii) where $\ltwo < x < \lone$ with $|x-L_i|\gg h_\mathrm{a}$ where  $h_\mathrm{a}$ represents the order of magnitude of the adsorption layers. There, the thickness of liquid~1 is large and the wetting energies $f_1$ and $f_3$ and their derivatives all approach zero. Furthermore, we assume the adsorption layer of liquid 2 on liquid 1 to be of uniform thickness $\hat{h}_\mathrm{a2}$, i.e., slopes and curvatures of the two  profiles are identical ($\partial_x\hontw=\partial_x\htwth$, $\kappa_{12}=\kappa_{23}$). 
Therefore, $\xiontw=\xitwth=1/\cos(\Phi_{13})$, where $\Phi_{13}$ is the angle of the local tangent of the curved interface with the horizontal. 
From \eqref{eq:mesoI_eula1} and \eqref{eq:mesoI_eula2} we then obtain
\begin{align}
&	\gontw \kappa_{12} + \xitwth f_2^\prime(\hhaB)=P_2-P_1
	\label{eq:meso_limit_ii_condition1}\\
&	\left(\gtwth + f_2(\hhaB)\right) \kappa_{23} -\xitwth f_2^\prime(\hhaB) = -P_2\,. \label{eq:meso_limit_ii_condition2}
\end{align}
Their summation gives the mesoscopic Laplace law valid in region (ii):
\begin{align}\label{eq:meso_lapl13}
	\left(\gontw + \gtwth + f_2(\hhaB)\right) \kappa_{23} = -P_1\,.
\end{align}
Similarly, Eq.~\eqref{eq:mesoI_hamiltonian} becomes 
\begin{align}\label{eq:meso_limit_ii_condition3}
	E %
	=& \gsl+ \cos(\Phi_{13})(\gontw+\gtwth+f_2(\hhaB))  - P_1(\hontw-\bar{h}_1) - P_2(\hhaB-\bar{h}_2)\,.
\end{align}
Note that as discussed below, in general, $\hhaB\neq \haB$.

Third, we consider the central region (iii) where $x<\ltwo$ with $|x-\ltwo|\gg h_\mathrm{a}$. There, the thicknesses of both layers are large and all wetting energies can be neglected. 
Furthermore, we write the metric factors as $\xiontw=1/\cos(\Phi_{12})$ and $\xitwth=1/\cos(\Phi_{23})$ where $\Phi_{12}$ and $\Phi_{23}$ are the {respective} angles of the local tangents of the curved interfaces $\hontw$ and $\htwth$ with the horizontal. 
From \eqref{eq:mesoI_eula1} and \eqref{eq:mesoI_eula2} we then obtain the Laplace laws
\begin{align}
&	\gontw \kappa_{12} =P_2-P_1\,, \label{eq:meso_limit_iii_condition1}\\
&	\gtwth \kappa_{23} = -P_2\,. \label{eq:meso_limit_iii_condition2}
\end{align}
Finally, Eq.~\eqref{eq:mesoI_hamiltonian} becomes 
\begin{align}\label{eq:meso_limit_iii_condition3}
	E = \gsl + \gontw\cos(\Phi_{12}) + \gtwth\cos(\Phi_{23})   -P_1(\hontw -\bar{h}_1) - P_2(\htwth-\hontw -\bar{h}_2)\,.
\end{align}
Having discussed the individual asymptotic regions (i), (ii) and (iii), we now use that $E,$ $P_1$ and $P_2$ are constant across the system, thereby connecting the previously obtained relations. In principle, this is possible for arbitrary macroscopic liquid volumes, i.e., for finite pressures $P_1$ and $P_2$. However, here we consider drop volumes that are sufficiently large to assume $P_1, P_2\to 0$. In other words, we consider ranges that mesoscopically are far outside the contact regions but macroscopically are close to contact. 
In consequence, Eqs.~\eqref{eq:meso_limit_ii_condition3} and \eqref{eq:meso_limit_iii_condition3} relate the angles that the interfaces at the macroscopic meeting point of regions (ii) and (iii) form with the horizontal by
\begin{align}
	\gontw\cos(\Psi_{12}) + \gtwth\cos(\Psi_{23}) = \cos(\Psi_{13})(\gontw+\gtwth+f_2(\hhaB)). \label{eq:mesoI_neumann_horiz}
\end{align}
This is the horizontal component of the mesoscopic Neumann law. %

The thicknesses $\haA$ and $\haB$ of the adsorption layers on the solid substrate far away from the sessile compound drop [in region (i)] are obtained from the two relations \eqref{eq:meso_limit_i_condition1} and \eqref{eq:meso_limit_i_condition2} with $P_1, P_2\to0$, i.e., from $f_1^\prime(\haA) +f_3^\prime(\haA+\haB)=0$ and $f_2^\prime(\haB)+f_3^\prime(\haA+\haB)=0$. Note that in general, $\haB$ differs from the thickness $\hhaB$ of the adsorption layer of liquid 2 on liquid 1 in region (ii). The latter is obtained from Eq.~\eqref{eq:meso_limit_ii_condition1} or \eqref{eq:meso_limit_ii_condition2}, i.e., in the considered limit  $f_2^\prime(\hhaB)=0$. 

To obtain the vertical component of the mesoscopic Neumann law we employ the mechanical {analogon} of the motion of particles and consider the generalized momenta $\frac{\partial\mathcal{L}}{\partial(\partial_x\tilde h_j)}=\pi_j$ in region (ii) at some $x>\ltwo$ and in region (iii) at some $x<\ltwo$. In particular, we use the conservation of the total generalized momentum $\pi_\text{tot} =\pi_1+\pi_2$ across the liquid-liquid-gas contact region at $x=\ltwo$. It actually corresponds to a completely inelastic collision of two {pseudo} particles: Arriving from region (iii) the particles approach {each other} in the contact region, interact inelastically due to their interaction potential (the wetting energy $f_2(h_{23}-h_{12})$) and continue as a bound state in region (ii). In this interpretation, Eq.~\eqref{eq:mesoI_neumann_horiz} represents energy conservation. Specifically, it describes how the generalized kinetic energy after the collision is related to the inner energy of the bound state. This is further discussed in section~\ref{sec:long-wave-meso}. 

Under the above assumptions ($P_1, P_2\to 0$) the total momentum in region (iii) is
\begin{align}
	\pi_\text{tot} 
	= \gontw\frac{\partial_x\hontw}{\xiontw} + \gtwth\frac{\partial_x\htwth}{\xitwth}
	= \gontw\sin(\Psi_{12}) - \gtwth\sin(\Psi_{23}).
\end{align}
In region (ii), one has $\partial_x\hontw=\partial_x\htwth$ and obtains
\begin{align}
	\pi_\text{tot} 
	&= \gontw\frac{\partial_x\hontw}{\xiontw} + \frac{\partial_x\htwth}{\xitwth} \left( \gtwth + f_2(\hhaB) \right)\nonumber\\
	&=-\gontw\sin(\Psi_{13}) -\left( \gtwth + f_2(\hhaB)\right)\sin(\Psi_{13})\,.
\end{align}
Taken together, the two relations give
\begin{align}\label{eq:mesoI_neumann_vert}
	\gontw\sin(\Psi_{12}) + \left( \gontw + \gtwth + f_2(\hhaB)\right)\sin(\Psi_{13}) = \gtwth\sin(\Psi_{23})\,,
\end{align}
corresponding to the vertical component of the mesoscopic Neumann law.

Next, we consider energy conservation between regions (ii) and (i), i.e., across the solid-liquid-gas contact region at $x=\lone$ (again for $P_1, P_2\to 0$). Then, Eqs.~(\ref{eq:meso_limit_i_condition3}) and (\ref{eq:meso_limit_ii_condition3}) give for the angle $\Theta_{13}$ that the straight interface in region (ii) forms with the horizontal 
\begin{align}
	\cos(\Theta_{13})(\gontw + \gtwth + f_2(\hhaB)) = \gontw + \gtwth + f_1(\haA)  +  f_2(\haB) + f_3(\haA + \haB)\label{eq:meso_younglaw13}\,,
\end{align}
which is the mesoscopic Young law for the three-phase contact of substrate, liquid 1 and gas. Note that within the mechanical {analogon}, it corresponds to a completely inelastic collision of a particle with a resting particle of infinite mass. The total inner energy after collision is given by $f_1+f_2+f_3$. Due to the infinite mass, the momentum of the impacting particle is simply absorbed, and no equation of momentum conservation {in the mechanical analogon} can be given.

In other configurations of the compound drop, the interface between liquid~1 and liquid~2, and the interface between liquid~2 and fluid~3 may both independently approach the substrate {(cf.~in Fig.~\ref{fig:sketch_macro_meso} the bottom row the inner and outer contact line, respectively)}, then similar considerations result in the mesoscopic Young laws
\begin{align}
	\gontw \cos \Theta_{12} = \gontw + f_1({\hat h}_{\mathrm{a}1})  \,,\label{eq:meso_younglaw12}
\end{align}
and
\begin{align}
	\gtwth \cos\Theta_{23} = \gtwth + f_1(\haA)  +  f_2(\haB) + f_3(\haA+\haB) - f_1({\hat h}_{\mathrm{a}1}) \label{eq:meso_younglaw23}\,,
\end{align}
respectively. Here, $\hat{h}_{\mathrm{a}1}$ corresponds to the adsorption layer thickness of liquid~1 under a thick layer of liquid~2. Similar to the case of the macroscopic description (Appendix~\ref{sec:statics-macro}) one may combine the three mesoscopic Young laws from Eqs.~\eqref{eq:meso_younglaw13}, \eqref{eq:meso_younglaw12} and \eqref{eq:meso_younglaw23} into
\begin{align}
	\gontw\cos(\Theta_{12}) + \gtwth\cos(\Theta_{23}) = \cos(\Theta_{13})(\gontw+\gtwth+f_2(\hhaB)) \label{eq:meso-young-combined}
\end{align}
that is of exactly the same form as the horizontal component of Neumann's law but with the three Young angles.

\subsection{Meso-macro consistency condition}
\label{sec:statics-consistency}
Before we embark on a detailed discussion of the wetting energy we first establish conditions for the consistency of the presented mesoscopic description with the macroscopic picture. For ease of comparison, Appendix~\ref{sec:statics-macro} briefly reviews the macroscopic energy, its minimization, and the resulting Laplace, Neumann and Young laws.

\begin{table}[h!]
	\begin{center}
          \caption{Conditions that ensure consistency between mesoscopic and macroscopic descriptions. Within the mesoscopic description, the wetting energies are shown in their general form (third column) as well as {in the specific form resulting from Eq.~\eqref{eq:wettpot-specific} (fourth column). Further details} are given in section~\ref{sec:wetting-energies}. 
		}
		\label{tb:mesomacro}
		\begin{tabular}{|p{0.18\textwidth}|>{\centering\arraybackslash}p{0.15\textwidth}|>{\centering\arraybackslash}p{0.28\textwidth}|>{\centering\arraybackslash}p{0.35\textwidth}|} \hline
			\textbf{interface energy} & \textbf{macroscopic} & \textbf{mesoscopic} & \textbf{mesoscopic}\\
			&&[general]&[specific with \eqref{eq:wettpot-specific} and \eqref{eq:static:ha1}-\eqref{eq:static:ha4}]\\
			\hline
			solid-liquid~1 & $\gamma_{\mathrm{s}1}$ & $\gamma_{\mathrm{s}1}$ & $\gamma_{\mathrm{s}1}$\\[0.5ex]
			solid-liquid~2 & $\gamma_{\mathrm{s}2}$ & $\gamma_{\mathrm{s}1}+\gamma_{12}+f_1(\hat h_{\mathrm{a}1})$& $ \gson + \gontw-\frac{3 A_1}{10 \haA^2}$\\[0.5ex]
			\multirow{2}{*}{solid-gas} & \multirow{2}{*}{$\gamma_{\mathrm{s}3}$} & $\gamma_{\mathrm{s}1}+\gamma_{12}+\gamma_{23}+f_1(\haA)$ \linebreak $ + f_2(\haB) + f_3(\haA+\haB) $& $\gson + \gontw + \gtwth-\frac{3 A_1}{10 \haA^2}$ \linebreak $ -\frac{3 A_2}{10 \haB^2}  -\frac{3 A_3}{10 (\haA+\haB)^2}$\\[0.5ex]
			liquid~1-liquid~2 & $\gamma_{12}$ & $\gamma_{12}$& $\gamma_{12}$\\[0.5ex]
			liquid~1-gas & $\gamma_{13}$ & $\gamma_{12}+\gamma_{23} +f_2(\hhaB)$& $  \gontw + \gtwth-\frac{3 A_2}{10 \haB^2}$\\[0.5ex]
			liquid~2-gas & $\gamma_{23}$ & $\gamma_{23}$& $\gamma_{23}$\\
			\hline
		\end{tabular}
	\end{center}
\end{table}

The macroscopic description is based on the six interface energies: liquid~1 to gas $\gamma_{13}$, liquid~2 to gas $\gamma_{23}$, liquid~1 to liquid~2 $\gamma_{12}$, liquid~1 to solid substrate $\gamma_{\mathrm{s}1}$, liquid~2 to substrate $\gamma_{\mathrm{s}2}$, and gas to substrate $\gamma_{\mathrm{s}3}$. %
In contrast, the mesoscopic description presented above is based on the same $\gamma_{23}$, $\gamma_{12}$, $\gamma_{\mathrm{s}1}$, and, additionally, the wetting energy. In other words, three of the macroscopic parameters are replaced in the mesoscopic description by the wetting energy that is a function of the two layer thicknesses.

Comparing the mesoscopic Laplace, Neumann, and Young laws [\eqref{eq:meso_lapl13}, \eqref{eq:meso_limit_iii_condition1}, \eqref{eq:meso_limit_iii_condition2}; \eqref{eq:mesoI_neumann_horiz}, \eqref{eq:mesoI_neumann_vert}; \eqref{eq:meso_younglaw13}, \eqref{eq:meso_younglaw12}, \eqref{eq:meso_younglaw23}] with their macroscopic equivalents [\eqref{eq:macro_laplace13}, \eqref{eq:macro_laplace12}, \eqref{eq:macro_laplace23}; \eqref{eq:macroI_neumannhoriz}, \eqref{eq:macroI_neumannvert}; \eqref{eq:macroI_younglaw13}, \eqref{eq:macroI_younglaw12}, \eqref{eq:macroI_younglaw23}, {respectively], one finds that full consistency of macroscopic and mesoscopic description is ensured by the equivalences listed in Table~\ref{tb:mesomacro}.} Next, we discuss more in detail the conditions this imposes on the wetting energy.

\subsection{Wetting energies}
\label{sec:wetting-energies}

When passing from the general gradient dynamics model in section~\ref{sec:model:dynamics} to the static considerations in section~\ref{sec:statics} we have in Eq.~\eqref{eq:gg} specified the general form of the wetting energy to be the sum of three contributions. 
{Thereby}, $f_1(\hontw)$ encodes the interaction of the bulk substrate with a bulk liquid 2 across a mesoscopic layer of liquid~1, $f_2(\htwth-\hontw)$ encodes the interaction of a bulk liquid~1 with the bulk gas phase across a mesoscopic layer of liquid~2, and $f_3(\htwth)$ represents a correction that ensures that the sum of all three terms encodes the interaction of the bulk substrate with the bulk gas phase across mesoscopic layers of liquid~1 and liquid~2. The derivation of the mesoscopic Neumann law has made it clear that $f_2$ has to be multiplied by the metric factor $\xitwth$ to correctly contribute to the macroscopic interface {energy} $\gamma_{13}$.

An obvious question is whether the term $f_3(\htwth)$ is indeed necessary. Pursuing this by setting $f_3=0$, i.e., by assuming that wetting interactions can be completely encoded by adding one contribution only related to the lower layer and another contribution only related to the upper layer, would directly result in the identifications $\hat h_{\mathrm{a}1}=\haA$ and $\hhaB=\haB$. This in turn would imply the relation $\gamma_{\mathrm{s}2}-\gamma_{12}=\gamma_{\mathrm{s}3}-\gamma_{13}$ between macroscopic quantities. As such a relation does, in general, not hold, the simplification {$f_3=0$} should not be employed. In other words, the contribution $f_3(h_{1}+h_{2})$ has to be incorporated for a faithful mapping of all macroscopic parameters {for partially wetting liquids} into the mesoscopic model. {Note that one is still able to construct a model} such that $\hat h_{\mathrm{a}1}=\haA$ and $\hhaB=\haB$ as {well} as $f_3(\haA+\haB)\neq0$. In terms of the general wetting energy $g(h_1, h_2)$ the condition is $g(\haA, \haB)\neq g(\hat h_{\mathrm{a}1}, \infty)+g(\infty,\hat h_{\mathrm{a}2})$. 

Overall, the formulation of the mesoscopic model either with layer thicknesses  $h_1$ and $h_2$ or interface height profiles  $h_{12}$ and $h_{23}$ together with the above discussion of the  three contributions to the wetting energy allows for a consistency check that should be applied to existing literature models: In the formulation with $h_1$ and $h_2$ one should have three terms $f_1(h_1)+f_2(h_2)+f_3(h_1+h_2)$ (here, for simplicity we dropped the metric factor). This implies that the variations of the energy w.r.t.\ $h_1$, and $h_2$ should contain contributions $f_1'(h_1)+f_3'(h_1+h_2)$, and $f_2'(h_2)+f_3'(h_1+h_2)$, respectively. In contrast, in the formulation with $h_{12}$ and $h_{23}$ one should have three terms $f_1(h_{12})+f_2(h_{23}-h_{12})+f_3(h_{23})$. Then, the variations of the energy w.r.t.\ $h_{12}$, and $h_{23}$ should contain contributions $f_1'(h_{12})-f_2'(h_{23}-h_{12})$, and $f_2'(h_{23}-h_{12})+f_3'(h_{23})$, respectively. Literature models that do not follow such a structure either unduly limit the macroscopic parameter space or break basic physical principles. 

For the example calculations shown below, we employ a specific thickness-dependence in $f_1$, $f_2$ and $f_3$. We combine a long-range van der Waals interaction with a short-range power law, i.e.,
\begin{align}
  f_i(h) = \frac{A_i}{2h^2} \left[\frac{2}{5}\left(\frac{h^\star_i}{h}\right)^3 - 1\right]
  \label{eq:wettpot-specific}
\end{align}
where $h$ stands for the relevant argument, the $A_i$ are Hamaker-type constants, and the $h^\star_i$ are parameters that represent typical length scales related to the equilibrium {adsorption layer thicknesses $h_{\mathrm{a}1}$, $h_{\mathrm{a}2}$, $\hat h_{\mathrm{a}1}$ and $\hat h_{\mathrm{a}2}$.} For $A_i>0$ the short-range interaction is repulsive ({here,} stabilizing) while the long-range one is attractive ({here,}, destabilizing) as appropriate for partially wetting liquids. The function $f_i(h)$ has a minimum $f_i(h^\star_i)=-3A_i/10 (h^\star_i)^2$ at $h=h^\star_i$, diverges {toward $\infty$} for $h\to0$, and approaches zero {from below} at large $h\to\infty$. {Note that this relatively simple choice is sufficient for our present purposes. However, to study more involved phenomena, e.g., involving first order wetting transitions, more complicated wetting energies should be employed, e.g., featuring a maximum at an intermediate thickness or several minima \cite{TeDS1988rpap,StVe2009jpm,HuTA2017jcp,YSTA2017pre}.}

Assuming flat interfaces and large drops ($P_1, P_2, \kappa_{12}\to 0$), one can determine the thicknesses of all coexisting adsorption layers by solving Eqs.~\eqref{eq:meso_limit_i_condition1}, \eqref{eq:meso_limit_i_condition2}, \eqref{eq:meso_limit_ii_condition1} and the equilibrium condition for an adsorption layer of liquid~1 underneath a bulk liquid~2, i.e.,  by solving the system   
\begin{align}
f_1^\prime(\haA) & = -f_3^\prime(\haA+\haB) \label{eq:static:ha1}\\
f_2^\prime(\haB)&= -f_3^\prime(\haA+\haB) \label{eq:static:ha2}\\
0&=f_2^\prime(\hhaB) \label{eq:static:ha3}\\  
0&=f_1^\prime(\hat{h}_{\mathrm{a}1}) \label{eq:static:ha4}
\end{align}
for $\haA$, $\haB$, $\hat{h}_{\mathrm{a}1}$ and $\hhaB$.  The specific form~\eqref{eq:wettpot-specific} with independent parameters $h^\star_i$ may be numerically solved. However, we can further simplify and still access the complete relevant macroscopic parameter space discussed in Table~\ref{tb:mesomacro}. Namely, we assume that $f_1$, $f_2$, and $f_3$ have their minima at  $\haA$, $\haB$, and $\haA+\haB$, respectively. This holds if in Eq.~\eqref{eq:wettpot-specific} one sets $h^\star_1=\haA$, $h^\star_2=\haB$ and $h^\star_3=\haA + \haB$, and further implies $\hat{h}_{\mathrm{a}1}=\haA$ and $\hhaB=\haB$. 

\begin{figure}[hbtp]
	\centering
	\includegraphics[width=0.9\textwidth]{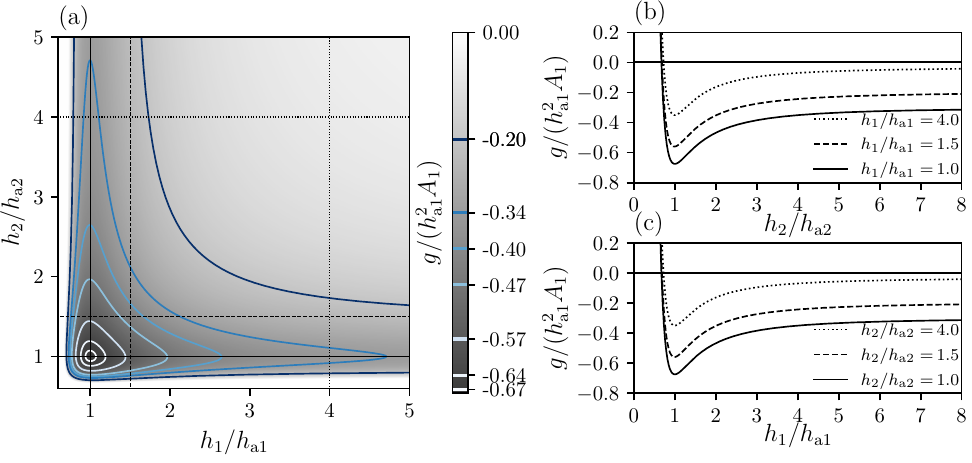}
	\caption{Shown is a typical wetting potential [\eqref{eq:gg} and \eqref{eq:wettpot-specific}] in the scaled form ${g}/(h_{\mathrm{a}1}^2 A_1)=[f_1(h_1)+f_2(h_2)+f_3(h_1,h_2)]/(h_{\mathrm{a}1}^2 A_1)$ with equal Hamaker constants and for homogeneous films, i.e., $\xi_{23}=1$. Panel~(a) gives the potential as a contour plot in the $(h_1,h_2)-$plane where positive values are suppressed for the sake of clarity. Specific values of chosen equipotential lines are given in the color bar. Panels~(b) and (c) present cuts at several fixed $h_1$ and $h_2$, respectively, as indicated in (a) by respective vertical and horizontal lines of corresponding line style. \label{fig:wetting_potential_withcuts}
	}
\end{figure}

A typical resulting wetting potential is presented in nondimensional form in Fig.~\ref{fig:wetting_potential_withcuts}. Note that expressions \eqref{eq:static:ha1} to \eqref{eq:static:ha4} only hold for $P_1, P_2\to 0$. In other words, the adsorption layer thicknesses slightly change for finite pressures. Normally, an increase in pressure results in a slight increase in the corresponding thickness. Geometrically, finite pressures within the two liquids (e.g., Laplace pressures within compound drops of finite volumes) result in a tilt of the energy in Fig.~\ref{fig:wetting_potential_withcuts} that slightly shifts the minima.

Introducing the resulting $f_i(h^\star_i)$ for $P_1, P_2\to 0$ into the expressions of column~3 of Table~\ref{tb:mesomacro} one obtains the consistency relations given in column~4 of Table~\ref{tb:mesomacro}. Vice versa, the three Hamaker constants $A_i$ can be directly related to the macroscopic interface energies by 
\begin{align}
	A_1 &=-\frac{10\haA^2}{3}(\gam{\mathrm{s}2} - \gam{\mathrm{s}1} - \gam{12}) \label{eq:HamakerconstantA1}\\
	A_2 &=-\frac{10\haB^2}{3}(\gam{13} - \gam{12} - \gam{23})\label{eq:HamakerconstantA2}\\
	A_3 &=-\frac{10 (\haA+\haB)^2}{3}(\gam{12} - \gam{13} + \gam{\mathrm{s}3} - \gam{\mathrm{s}2})\,. \label{eq:HamakerconstantA3}
\end{align}
As in the present setup, the Hamaker constants have to be positive to ensure short-range repelling and long-range attracting interactions between interfaces, Eqs.~\eqref{eq:HamakerconstantA1}-\eqref{eq:HamakerconstantA3} imply three inequalities that the macroscopic interface energies have to fulfill to allow for a mesoscopic description by a wetting energy that combines expressions of the form \eqref{eq:wettpot-specific}. {For instance, Eq.~\eqref{eq:HamakerconstantA2} shows that $A_2>0$ directly corresponds to the macroscopic Neumann triangle inequality $\gam{12} + \gam{23}>\gam{13}$.}

Furthermore, the sequence of the two liquids has to be chosen as energetically favorable, i.e., $\gamma_{\mathrm{s}1} + \gontw + \gtwth < \gamma_{\mathrm{s}2} + \gontw + \gonth$, therefore, a further inequality to hold is $\gamma_{\mathrm{s}1} - \gamma_{\mathrm{s}2}< \gonth - \gtwth$. Note, that \eqref{eq:HamakerconstantA3} well illustrates that neglecting $f_3$ (i.e., $A_3=0$) is only valid for the particular values that fulfill $\gam{\mathrm{s}2} - \gam{12} =  \gam{\mathrm{s}3} - \gam{13}$, {i.e., represents a nongeneric case.}

{In summary, the restrictions of the considered parameter space, on the one hand, result from our focus on the case of partial wettability. On the other hand, they result from our mesoscopic modeling approach where, e.g., all contact angles need to be smaller than $\pi/2$, and situations like encapsulation of one of the liquids inside a drop of the other liquid are not considered.}

Up to here, all relations are derived within the full-curvature variant {of the mesoscopic model.} In the next section, we consider how the various obtained laws and relations simplify when considering {the long-wave variant}.

\subsection{Long-wave approximation}
\label{sec:long-wave-meso}

To apply a long-wave approximation \cite{OrDB1997rmp,CrMa2009rmp} one assumes that all relevant length scales parallel to the substrate are large compared to relevant vertical scales. 
This implies that all interface slopes, in particular, Young angles and Neumann angles w.r.t.\ the horizontal, have to be small. In the energy functional \eqref{eq:meso_grandpotential} the approximated metric factors are  $\xi^{\mathrm{lw}}_{12}=1+\frac{1}{2}|\nabla h_{12}|^2$ and  $\xi^{\mathrm{lw}}_{23}=1+\frac{1}{2}|\nabla h_{23}|^2$ resulting as before in the variations \eqref{eq:mesoI_eula1} and \eqref{eq:mesoI_eula2} but with approximated curvatures $\kappa^{\mathrm{lw}}_{12}= \partial_{xx} h_{12}$ and $\kappa^{\mathrm{lw}}_{23}= \partial_{xx} h_{23}$. 

Considering as above the three regions (i), (ii) and (iii) (cf.~Fig.~\ref{fig:sketch_macro_meso}) and subsequently the conservation of energy and generalized momentum in the mechanical {analogon}, we obtain the two components of the long-wave Neumann law
\begin{align}
\gontw\Psi_{12} + \left( \gontw + \gtwth + f_2(\hhaB)\right)\Psi_{13} &= \gtwth\Psi_{23}\label{eq:mesoI_neumann_vert-lw} \\
\gontw \Psi_{12}^2 + \gtwth \Psi_{23}^2 -(\gontw+\gtwth+f_2(\hhaB))\Psi_{13}^2&= -2 f_2(\hhaB)\label{eq:mesoI_neumann_horiz-lw}
\end{align}
and the three Young laws
\begin{align}
(\gontw + \gtwth + f_2(\hhaB)) \Theta_{13}^2 &= -2 ( f_1(\haA)  +  f_2(\haB) + f_3(\haA + \haB) - f_2(\hhaB)) \label{eq:mesoI_younglaw13-lw}\\
\gtwth \Theta_{23}^2 &=  -2( f_1(\haA)  +  f_2(\haB) + f_3(\haA + \haB) - f_1({\hat h}_{\mathrm{a}1})) \label{eq:mesoI_younglaw23-lw}\\
	\gontw \Theta_{12}^2 &= -2f_1({\hat h}_{\mathrm{a}1})  \label{eq:meso_young12-lw}
	.
\end{align}
Using the consistency conditions in Table~\ref{tb:mesomacro}, Eqs.~\eqref{eq:mesoI_neumann_vert-lw} to \eqref{eq:meso_young12-lw} are fully consistent with the long-wave variant \eqref{eq:macroI_neumannvert-lw} to \eqref{eq:macroI_younglaw12-lw}, respectively,  of the macroscopic laws in Appendix~\ref{sec:statics-macro}.

Note that the mechanical {analogon} discussed above in the context of the vertical component of Neumann's law, {in the long-wave case} corresponds to standard Newtonian mechanics of two point particles: For $P_1, P_2\to 0$, Eqs.~\eqref{eq:mesoI_neumann_horiz-lw} and \eqref{eq:mesoI_neumann_vert-lw} correspond to energy and momentum conservation during a collision of point particles at positions $h_{12}$ and $h_{23}$ with interaction potential {$f_2(h_{23}-h_{12})$, see above.} The interface energies $\gam{12}$ and $\gam{23}$ {then} represent the corresponding inertial masses. Incorporating finite pressures into the relations would amount to adding an external gravitational potential with the respective gravitational masses being proportional to $P_1$ and $P_2$. 

Our discussion of the wetting energy $g$ has made it clear that the Derjaguin (or disjoining) pressures that enter the pressure expressions for the two layers are obtained as derivatives of the total wetting energy w.r.t.\ the relevant layer heights or thicknesses (depending on the particular formulation, see section~\ref{sec:model:dynamics}).\footnote{If this is not taken into account Newton's third law is broken. In our opinion this occurs in Ref.~\cite{YaKK2012prb} (their Eqs.~(3)-(5) are inconsistent, possibly due to a missing term in Eq.~(3)).} Also note that none of the literature models cited in sections~\ref{sec:model:dynamics} and~\ref{sec:statics} takes the metric factor in front of $f_2(h_{23}-h_{12})$ into account, {cf.~Eq.~\eqref{eq:meso_grandpotential}}. In consequence, to our knowledge, none of the existing mesoscopic two-layer models provides exact mesoscopic Neumann and Young laws {either within the full-curvature variant or the long-wave variant} that are consistent with the corresponding macroscopic laws.
\section{Applications}
\label{sec:appl}

In the previous section, we have introduced a mesoscopic hydrodynamic model for a two-layer system of two nonvolatile immiscible partially wetting liquids. The employed gradient dynamics framework ensures thermodynamic consistency. Furthermore, consistency with the macroscopic description of two-liquid compound drops is guaranteed by three conditions formulated for the wetting energy that relate macroscopic parameters to mesoscopic ones. Additionally, the vertical order of layering implies a fourth condition in the form of an inequality that connects four of the six macroscopic interface energies. Next, we apply the established model to a number of typical situations and discuss selected {results obtained by direct numerical simulations of the dynamic model~\eqref{eq:pde_system}.} We remind the reader that one may use either of the two given mesoscopic formulations, i.e., in terms of layer thicknesses (Eq.~\eqref{eq:pde_system}-\eqref{eq:mobilitymatrix}) or in terms of interface heights (Eq.~\eqref{eq:trans-gen} and~\eqref{eq:mobilitymatrix-alt}). Both exist as full-curvature and long-wave variants. Here, we employ the full-curvature variant and effectively nondimensionalize the system by setting $\gontw=\eta_1=\haA=1$.  Additionally, we set the interface energy between liquid~1 and the substrate to $\gson=0 $ as only differences in substrate energies enter (macroscopic) Young laws. All numerical results are obtained by using the finite-element library \textit{oomph-lib} providing space- and time-adaptivity \cite{HeHa2006}.
\subsection{Spreading compound drops}
\label{sec:spreading}

\begin{figure}[bth]
	\includegraphics[width=0.9\textwidth]{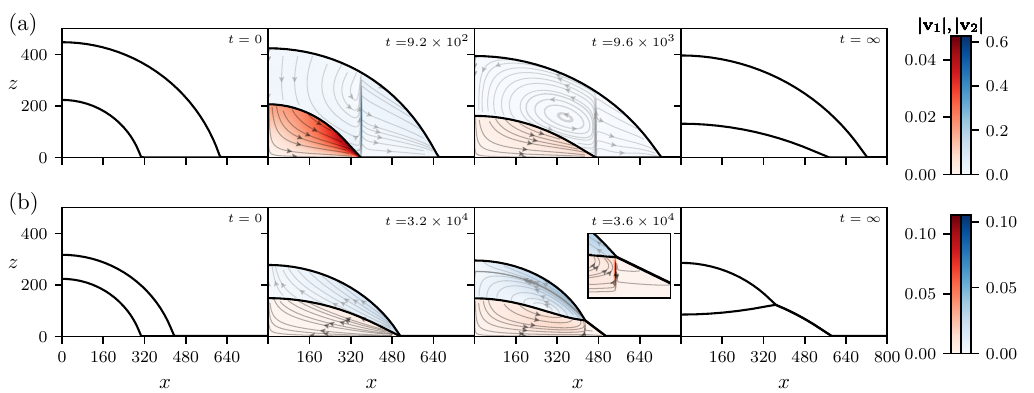}
	\caption{
	Shown are snapshots of the relaxational spreading dynamics of a compound drop for (a) $V_2/V_1=3$ and (b) $V_2/V_1=1$. In both cases, an initial drop-covers-drop configuration is used (left-most panels), i.e.,  a drop of liquid~2 entirely covers a drop of liquid~1. As reflection symmetry holds during the entire evolution, only the right hand side of the profiles is shown. Streamlines indicate velocity fields. Additionally, their absolute values are shown by colored shading. Further details and parameter values are given in the main text, further analysis is provided in Fig.~\ref{fig:spread_diss_en_st}. 
	Also, consider the corresponding movies (file numbers 00 and 01) in the Supplementary Material \cite{DiTh2024suppmat}. 
	}
	\label{fig:sim_spread}
\end{figure}

\begin{figure}[tbh]
	\includegraphics[width=0.9\textwidth]{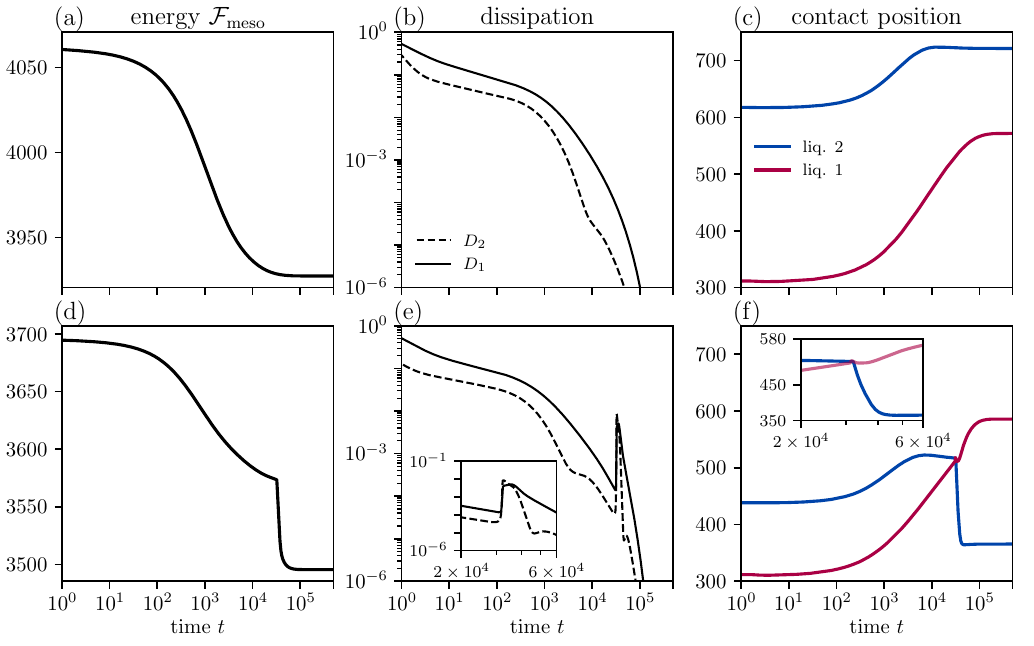}
	\caption{The top and bottom row provide further analysis of the dynamics shown in Figs.~\ref{fig:sim_spread}~(a) and \ref{fig:sim_spread}~(b), respectively.
    Given are the time dependencies of [(a),(d)] the free energy [Eq.~\eqref{eq:energy-meso}], [(b),(e)] the dissipation rates in liquid~1 ($D_1$) and liquid~2 ($D_2$), (cf.~section~\ref{sec:app:supp}), and [(c),(f)] the contact line positions. The insets in panels (e)~and~(f) magnify the time interval when the two contact line regions cross.
	}
	\label{fig:spread_diss_en_st}
\end{figure}

\begin{figure}[tbh]
	\includegraphics[width=0.9\textwidth]{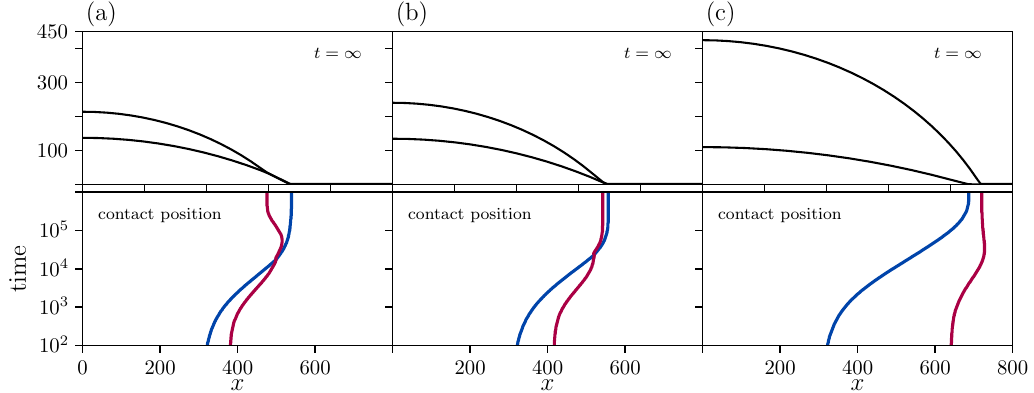}
	\caption{
		Shown are equilibrated compound drops together with corresponding space-time diagrams of the contact line positions where the liquid 1-liquid 2 or the liquid 1-gas interface meet the
substrate (blue line) and where the liquid 2-gas interface meets the substrate or the
three fluids meet (red line). The volume ratios $V_2/V_1$ are (a) 0.5, (b) 0.8, and (c) 3.3. The interface energies are $(\gonth, \gtwth, \gstw, \gsth)=(1.3,0.31,0.95,1.1)$ while all other parameters are as in Fig.~\ref{fig:sim_spread}. 
	}
	\label{fig:spread_four_phase}
\end{figure}

First, we numerically investigate how sessile compound drops spread. We employ a symmetric initial configuration, namely, a drop of liquid~1 that is entirely covered by a drop of liquid~2, i.e., a \textit{drop-covers-drop} configuration. We consider a one-dimensional substrate, i.e., the liquid-liquid and liquid-gas interfaces are circular arcs. The volume of liquid~1 is fixed at $V_1=10^5$ and two different values for the volume $V_2$ of liquid~2 are used. {Note that these volumes are defined as the excess volume above the adsorption layers, e.g., $V_1=\int(h_1-h_\mathrm{a1})\mathrm{d}x$.} The dynamic viscosity of liquid~2 is $\eta_2/\eta_1=0.8$ and its adsorption layer height is $\haB/\haA=1$.  The interface energies are set to $(\gonth, \gtwth, \gstw, \gsth)=(2,1.3,0.9,1.6)$ such that all four inequalities discussed at the end of Section~\ref{sec:wetting-energies} hold.  The resulting macroscopic equilibrium contact angles are $(\theta_{12},\theta_{13},\theta_{23})\approx(25.8\,^\circ,36.9\,^\circ,57.4\,^\circ)$.  As the initial droplets have contact angles $\theta_{12}^\mathrm{init}=\theta_{23}^\mathrm{init}=72\,^\circ$ larger than the corresponding equilibrium values both liquids spread.  Snapshots from the simulations of the spreading processes are shown in Fig.~\ref{fig:sim_spread}~(a) and (b) for the volume ratios $V_2/V_1=3$ and $V_2/V_1=1$, respectively.

For the compound drop of large volume ratio, the configuration does not change qualitatively during the relaxational spreading process, see Fig.~\ref{fig:sim_spread}(a). Both liquids spread till the contact angles have reached their equilibrium values, and always remain in drop-covers-drop configuration. A deeper insight into the dynamics of the deformable interfaces is attained by inspecting the vertically resolved velocity fields within the two liquids that are reconstructed from the film thickness profiles as detailed in appendix~\ref{app:velocity_fields}. These fields are incorporated in the respective two central panels of Figs.~\ref{fig:sim_spread}~(a) and (b) as selected streamlines and coloring that encodes the absolute value of the local velocity. The larger velocities in the vicinity of the contact lines result from strong changes in pressure gradients (cf.~\ref{app:velocity_fields}).\footnote{Due to the limited exactness of the plotting routines the coloring should be interpreted as providing the relative strength of the local flow.}  

In contrast, at the smaller volume ratio employed in Fig.~\ref{fig:sim_spread}~(b) the configuration changes during relaxation.  As in the previous case, both liquids initially spread, however, at $t\approx 3.2\cdot 10^{4}$ the contact line where the liquid~1-liquid~2 interface meets the substrate catches up with the one of the liquid~2-gas interface. In consequence, the substrate-liquid~2 interface shrinks to zero and a transient four-phase contact is formed. Afterward, the liquid~1-liquid~2-gas contact separates from the substrate and moves inwards, i.e., a drop of liquid~2 retracts onto the drop of liquid~1 (that still spreads) until equilibrium Neumann angles are reached. In the final state, liquid~2 forms a drop positioned symmetrically on top of a drop of liquid~1, i.e., a \textit{drop-on-drop} configuration. During the spreading process the liquid flow can either form a convection roll-like structure in one of the liquids or across both liquids as in the third panels of Figs.~\ref{fig:sim_spread}~(a) and \ref{fig:sim_spread}~(b) or be predominantly parallel as in the second panel of Fig.~\ref{fig:sim_spread}~(b) when both drops spread at similar speeds. Note that the nearly vertical structure visible in some panels at the location of the inner contact line reflects that, e.g., a faster spreading inner drop can drag the upper liquid along and \enquote{push it} into the more quiescent outer liquid in front of the advancing inner contact line. A magnification of such a region is shown in the inset of the third panel of  Fig.~\ref{fig:sim_spread}~(b). {We point out that simulations with the long-wave variant of the model at identical parameters and initial conditions as Fig.~\ref{fig:sim_spread} (not shown) result in identical qualitative behavior as obtained with the full-curvature variant.
  In particular, the equilibrated compound drops show the identical configurations in both variants. Instead of circular arcs in the full-curvature variant the final height profiles in the long-wave variant are composed by parabolas. Only pressure gradients and, therefore, driving forces and velocities quantitatively differ for the two variants.}

Further quantitative information about the dynamic processes showcased in Fig.~\ref{fig:sim_spread}  can be found in Fig.~\ref{fig:spread_diss_en_st}~(a)-(c) for $V_2/V_1=3$, and Fig.~\ref{fig:spread_diss_en_st} ~(d)-(f) for $V_2/V_1=1$. Thereby, the first column displays the dependence of the free energy~\eqref{eq:energy-meso} on time. When both drops continuously spread keeping the initial drop-covers-drop configuration [Fig.~\ref{fig:spread_diss_en_st}~(a)] the energy decreases rather smoothly till it approaches its equilibrium values. In the process, interface energy is viscously dissipated within liquid~1 (rate $D_1$) and liquid~2 (rate $D_2$), see Fig.~\ref{fig:spread_diss_en_st}~(b). The dissipation rates are determined using Eq.~\eqref{eq:diss_1} that is derived in Appendix~\ref{app:dissipation}. 
As the viscous dissipation in the vicinity of a  contact line dominates (cf.~Ref.~\cite{EWGT2016prf}) and is larger for smaller contact angles, in the considered case of similar viscosities the always smaller  
contact angle ($\theta_{12}<\theta_{23}$) at and the larger velocity of the inner contact line implies $D_1>D_2$ for the entire process as observed in Fig.~\ref{fig:spread_diss_en_st}~(b).
Finally, the third column of Fig.~\ref{fig:spread_diss_en_st} characterizes the overall dynamics by showing the two contact line positions over time. It clearly indicates that the configuration does not change (no change in sequence and/or number of contacts), that the inner contact line is always faster than the outer one (steeper slope of the red line), and that the outer contact angle is the first to reach equilibrium. The slight overshoot of the outer contact line is a dynamic effect of the still spreading inner drop.

In contrast to Fig.~\ref{fig:spread_diss_en_st}~(a), in the case of Fig.~\ref{fig:spread_diss_en_st}~(d) the energy seems to converge at $t\approx 10^4$ but then abruptly decreases in a nearly step-wise manner. This is related to the approach of the two contact lines (still as drop-covers-drop configuration) into the short-lived  transient of four-phase coexistence and the subsequent configuration change to drop-on-drop {configuration}, i.e., the outer contact line has transformed into an inner (dynamic) Neumann region. {This} dramatic change is also reflected in features of the time dependencies of dissipation rates depicted in Fig.~\ref{fig:spread_diss_en_st}~(e). Both show a local minimum at {the transient} four-phase coexistence where contact regions \enquote{cross} and their speeds are minimal, see Fig.~\ref{fig:spread_diss_en_st}~(f). Afterwards, $D_2$ and $D_1$ first sharply rise before decreasing again till finally equilibrium is reached. Close inspection of Fig.~\ref{fig:spread_diss_en_st}~(f) shows {that before contact regions cross at $t\approx10^4$} the outer (liquid~2-gas) contact line slightly recedes. {Then, directly} after the crossing the same happens for the new outer (liquid~1-gas) contact line. {This indicates} a direct attractive interaction between contact regions. {As the four-phase coexistence only exists for a very brief transient time span it most likely does not correspond to a metastable state (a local minimum of the energy). However, the seemingly convergent energy shortly before the step-change indicates that it might correspond to an unstable steady state representing a saddle point of the energy. Such a saddle is approached by the trajectory of the system along a stable direction in phase space before it is left behind when the trajectory is captured by an unstable direction, and ultimately converges to a stable state.}

We further scrutinize the occurrence of four-phase contact by considering a different set of interface energies where, starting with initial conditions as in Fig.~\ref{fig:sim_spread}, for a range of volume ratios the compound drop may actually relax toward a \textit{four-phase-contact} configuration. In particular, for $(\gonth, \gtwth, \gstw, \gsth)=(1.3,0.31,0.95,1.1)$ (resulting in equilibrium contact angles $(\theta_{12},\theta_{13},\theta_{23})\approx(18.2\,^\circ,32.2\,^\circ,61.1\,^\circ)$), we find the equilibrated compound drops shown in the top row of Fig.~\ref{fig:spread_four_phase}, also see the corresponding space-time plots of the contact lines positions in the bottom row. For $V_2/V_1=0.5$ [Fig.~\ref{fig:spread_four_phase}(a)] and $V_2/V_1=3.3$ [Fig.~\ref{fig:spread_four_phase}(c)] the resulting equilibrium configurations are similar to those of Fig.~\ref{fig:sim_spread}(b) and (a), respectively, i.e., drop-on-drop and drop-covers-drop. However, in contrast to Fig.~\ref{fig:sim_spread}(b), in the case of Fig.~\ref{fig:spread_four_phase}(a) there is a finite time-span with four-phase contact where both contact regions move together. Subsequently, they separate again, the newly formed Neumann contact line region moves inwards while the liquid 1-gas-substrate contact line continues to move outwards before both reach their equilibrium positions in the drop-on-drop configuration.
One may say that the motion, interaction and crossing of the two contact line regions parallels certain aspects of soliton-soliton interactions \cite{InfeldRowlands1990}.

However, in a small range of intermediate volume ratios, e.g., at $V_2/V_1=0.8$ shown in Fig.~\ref{fig:spread_four_phase}~(b), the contact line regions do not separate again and the four-phase-contact configuration corresponds to the equilibrium compound drop. The corresponding space-time plot reveals that this result is not just a coincidence for one specific volume ratio because the four-phase-contact is formed at a smaller $x$-value than the one of the final state, i.e., the combined contact region continues to move outwards towards the equilibrium state. The possibility of such an equilibrium configuration with four-phase contact that exists for a finite range of volume ratios is predicted in Ref.~\cite{MaAP2002jfm} based on macroscopic considerations as a state between drop-covers-drop and drop-on-drop configuration. This prediction is here confirmed employing a dynamic mesoscopic hydrodynamic model. However, it remains a task for the future to identify the corresponding parameter ranges where the various states are stable, metastable or unstable steady states of the model. 
\subsection{Sliding compound drop}
\label{sec:sliding}

\begin{figure}[tbh]
	\includegraphics[width=0.9\textwidth]{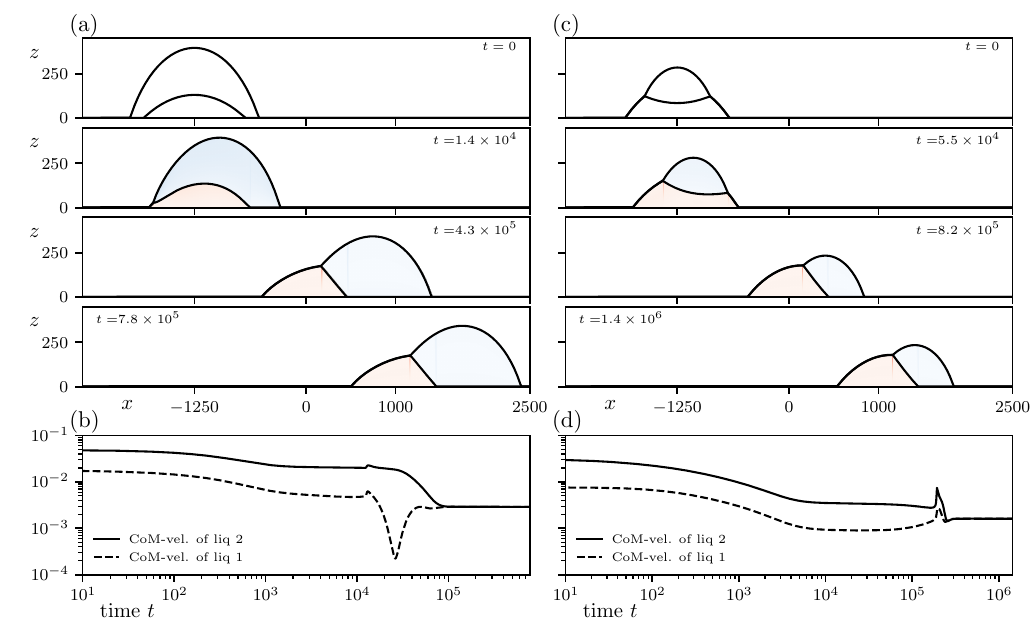}
	\caption{Time evolution of sliding compound drops: Equilibrium compound drops with (a)-(b) $V_2/V_1=3$ and (c)-(d) $V_2/V_1=1$ obtained in section~\ref{sec:spreading} are placed on an inclined plane with $\beta=1\cdot10^{-6}$. The respective upper four panels (in (a) and (c)) show snapshots at selected times while (b) and (d) show a log-log representation of the time dependence of the center-of-mass (CoM) velocity individually for the two liquids.  For further details see main text. Other parameters and shading are as in Fig.~\ref{fig:sim_spread}. Also consider the movies (file numbers 02 and 03) in the Supplementary Material \cite{DiTh2024suppmat}.} %
	\label{fig:sim_sliding}
\end{figure}

In the previous section, we have considered the spreading of compound drops on horizontal substrates and discussed their relaxational dynamics toward various equilibrium configurations.
Next, we investigate sliding compound drops on a homogeneous inclined substrate. As initial conditions we use the equilibrium drops obtained before. Thereby, we incorporate gravity while still considering small droplets. To focus on the sliding motion we neglect the hydrostatic pressure and only account for the symmetry-breaking downhill force parallel to the substrate.  We introduce it into the free energy $\mathcal{F}_\mathrm{meso}$ [Eq.~\eqref{eq:meso_grandpotential}] in analogy to Refs.~\cite{GTLT2014prl, HDGT2024l}. In the interface height formulation, the energy reads
\begin{align}\label{eq:free_energy_sliding}
 \widetilde{\mathcal{F}}_\mathrm{meso} = \mathcal{F}_\mathrm{meso} + \int_{-\infty}^\infty \beta x \htwth\mathrm{d}x\,,
\end{align}
with the inclination parameter $\beta=\rho g \sin\alpha$, where $\rho$ is the common mass density of the liquids, $g$ is the gravitational acceleration, and $\alpha$ is the inclination angle. Note that a generalization for different mass densities of the two liquids is straightforward but not pursued here. Inserting $\widetilde{\mathcal{F}}_\mathrm{meso} $ into  Eq.~\eqref{eq:trans-gen} the right-hand sides of the two dynamic equation acquire additional terms $-\beta\partial_x(\tilde{Q}_{1 2})$ and $-\beta\partial_x(\tilde{Q}_{2 2})$, respectively. 

Figs.~\ref{fig:sim_sliding}~(a) and (c) show snapshots of resulting time simulations for the volume ratios $V_2/V_1=3$ and $V_2/V_1=1$, respectively.  The inclination parameter is $\beta=10^{-6}$ while all other parameters are as in Fig.~\ref{fig:sim_spread}.  Initially, at $t=0$, the center-of-mass (CoM) of the equilibrated compound drops are positioned at $x=-1250$ within a computational domain of total size $L=5000$.  The respective next three panels display profiles at later times. Thereby the final two show converged stationary sliding states at times when the CoM of liquid~1 is at $x=0$ and $x=1000$, respectively. Fig.~\ref {fig:sim_sliding}~(b) and (d) give the CoM-velocities as a function of time in a log-log representation.

Inspecting the time evolution in Fig.~\ref{fig:sim_sliding}~(a) one sees that the initial symmetric drop-covers-drop configuration at $V_2/V_1=3$ first becomes asymmetric under the influence of the lateral driving force (also see movie with file number 02 in the Supplementary Material \cite{DiTh2024suppmat}). In other words the smaller inner drop slides more slowly than the outer drop. As a result, their trailing contact regions approach till the liquid~2-gas-substrate contact has caught up with the liquid~1-liquid~2-substrate contact at $t\approx1.2\cdot 10^4$. Then, the configuration changes into a strongly asymmetric \textit{drop-beside-drop} configuration, i.e., the trajectories of the contact regions cross. In the process, the  liquid~1-liquid~2-substrate contact becomes the trailing liquid~1-gas-substrate contact while the liquid~2-gas-substrate contact becomes a liquid~2-liquid-1-gas contact, i.e., a dynamic Neumann region. In parallel, the trailing liquid 2-substrate interface vanishes and a liquid~1-gas interface emerges. This transition results in a small peak of the CoM-velocities. In the subsequent time span the drop of liquid~1 comes nearly to rest with a minimal CoM-velocity of $v_1\approx2\cdot10^{-4}$ at $t\approx2.6\cdot 10^4$ while the drop of liquid~2 slides over it till the compound drop reaches the converged stationary sliding state that has a velocity of approximately $2.9\cdot 10^{-3}$ as obtained by a linear fit for the time interval bounded by liquid~1's CoM passing $x=0$ and $x=1000$, and by averaging the CoM-velocities of the two liquids.

Inspecting next the time evolution in Fig.~\ref{fig:sim_sliding}~(c) one notices that also the initial symmetric drop-on-drop configuration at $V_2/V_1=1$ becomes asymmetric, namely, the upper drop slides faster as it is \enquote{lubricated} by the lower drop. Ultimately, also here, the configuration changes into a stationary sliding asymmetric compound drop in drop-beside-drop configuration. In contrast to the previous case, now the crossing of contact regions occurs at the front end, i.e., the  liquid~1-liquid~2 interface becomes strongly tilted, a liquid~1-liquid~2-substrate contact emerges,  the leading liquid~1-liquid~2-gas becomes a liquid~2-gas-substrate contact, and a liquid 2-substrate interface emerges.  The process results in a local maximum of both CoM velocities at $t\approx 2.0\cdot 10^{5}$. The converged stationary sliding drop has a velocity of $1.6\cdot 10^{-3}$, i.e., it is slower than the one in Fig.~\ref{fig:sim_sliding}~(a).

\FloatBarrier
\subsection{Coarsening ensemble of drops}\label{sec:coarsening}

\begin{figure}[tbh]
	\includegraphics[width=0.9\textwidth]{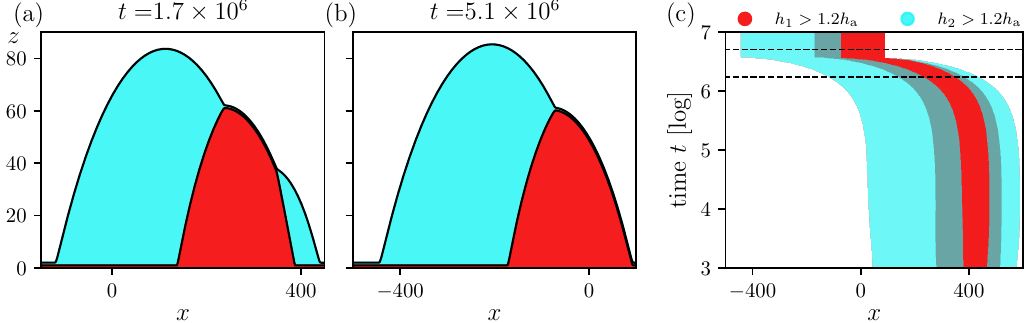}
	\caption{
		Shown is a minimal example of coarsening dynamics starting with a compound drop consisting of three subdrops (two of liquid~2 beside one of liquid~1). Panels (a) and (b) show selected snapshots of the profiles at times $t$ marked by dashed horizontal lines in the space-time plot of panel (c). The latter is given in log-normal form with liquid~1 and liquid~2 indicated by red and transparent blue color to visualize relative positions and layering. Only thicknesses above the threshold value of $1.2$ are shown. The  interface energies are $(\gonth, \gtwth, \gstw, \gsth)=(1.8,0.9,0.7,1.4)$, and the domain size is $L=1600$, note that this is only partly shown.
	}
	\label{fig:sim_coars_1d_1drop}
\end{figure}

Having discussed individual compound drops on horizontal and inclined homogeneous substrates, next, we consider the coarsening behavior of drop ensembles on horizontal substrates. We set the interface energies to $(\gonth, \gtwth, \gstw, \gsth)=(1.8,0.9,0.7,1.4)$ while the remaining parameters are as in section~\ref{sec:spreading}. 
The resulting equilibrium contact angles are $(\theta_{12},\theta_{13},\theta_{23})\approx(45.1\,^\circ,38.9\,^\circ,38.9\,^\circ)$. 

First, we consider a minimal example of coarsening, namely, a single initial compound drop consisting of a large and a small volume of liquid~2 to the left and right of a single intermediate-size volume of liquid~1, respectively, i.e., the compound drop consists of three subdrops, see Fig.~\ref{fig:sim_coars_1d_1drop}. This is similar to states visible in Fig.~8~(d) of Ref.~\cite{DGGR2021l} or the lower panel in Fig.~23 of Ref.~\cite{ATTG2024pre} where the phase decomposition of sessile drops of a mixture is considered.
Here, coarsening occurs as expected based on the greater Laplace pressure in the smaller subdrop, i.e., liquid~2 is transported from the right to the left via the adsorption layer of liquid~2 on liquid~1. During this process of Oswald ripening (coarsening by mass transfer), interestingly, the whole compound drop moves by nearly one drop base length to the left. This is well visible in the space-time plot (log-normal representation) in Fig.~\ref{fig:sim_coars_1d_1drop}~(c). The resulting equilibrated compound drop corresponds to an asymmetric \textit{drop-beside-drop} configuration similar to the states in the top and central panels in Fig.~23 of Ref.~\cite{ATTG2024pre}.

\begin{figure}[tbh]
	\includegraphics[width=0.9\textwidth]{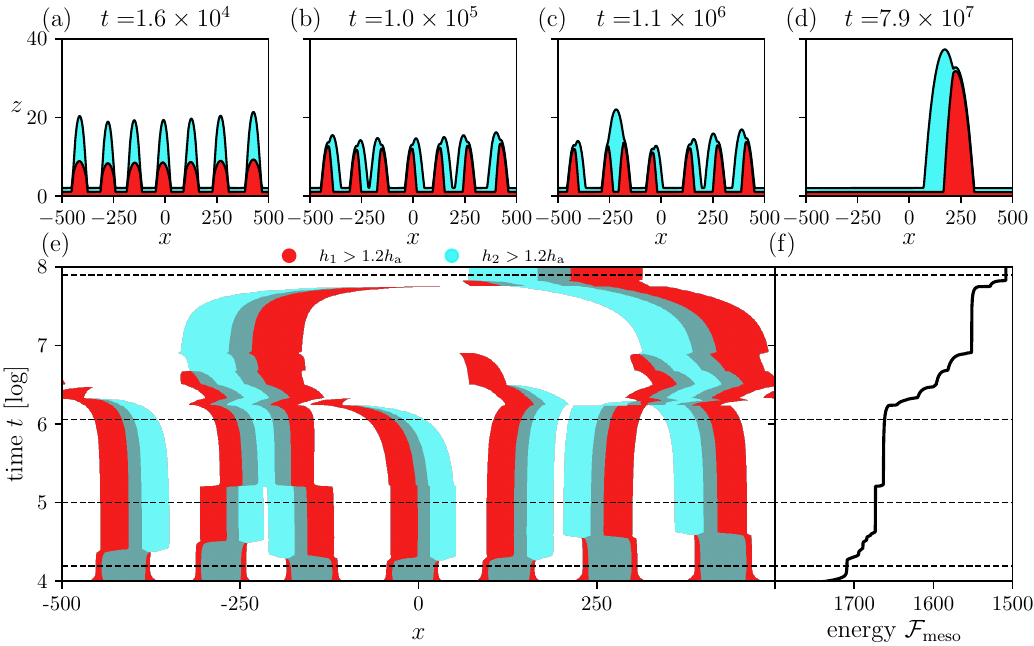}
	\caption{
          Shown is the coarsening dynamics of a small ensemble of compound drops. Panels~(a)-(d) show selected snapshots at times $t$ marked by dashed horizontal lines in the log-normal space-time plot of panel (e), and panel (f) gives  the corresponding dependence of the total  free energy on time. The mean thicknesses of the liquid layers are $\bar h_{1}=4$ and $\bar h_{2}=4$  and the domain size is $L=1000$. All remaining parameters are as in Fig.~\ref{fig:sim_coars_1d_1drop}. Also, consider the movie (file number 04) in the Supplementary Material \cite{DiTh2024suppmat}.
	}
	\label{fig:sim_coars_1d_ensemble}
\end{figure}

Next, we consider a larger ensemble of compound drops: Two initially homogeneous layers of thicknesses $\bar{h}_{1}=\bar{h}_{2}=4$ that are linearly unstable with respect to dewetting are perturbed by noise of small amplitude.  This results in spinodal dewetting as shown in Fig.~\ref{fig:sim_coars_1d_ensemble} similar to, e.g., \cite{PBMT2004pre,PBMT2005jcp,FiGo2005jcis,BaGS2005iecr,BaSh2006jcp}. Initially, for the chosen domain size fluctuations grow into seven compound drops. Each is in a nearly symmetric drop-on-drop configuration, see Fig.~\ref{fig:sim_coars_1d_ensemble}~(a). Interestingly, before drops coarsen, each individual drop undergoes a morphological transition from the symmetric drop-on-drop configuration to an asymmetric drop-beside-drop configuration, see Fig.~\ref{fig:sim_coars_1d_ensemble}~(b). This is unlike the processes shown in  \cite{PBMT2005jcp,BaSh2006jcp}. Subsequently, coarsening sets in and is normally accompanied by the above discussed lateral motion, see the corresponding space-time plot in Fig.~\ref{fig:sim_coars_1d_ensemble}~(e). Indeed, it shows that coarsening by translation is dominant while coarsening by mass transfer only scarcely occurs at later times. Translation seems to be of greater importance than in the coarsening of drops of a single simple liquid on rigid solid substrates \cite{GlWi2005pd,OtRS2006sjma,PCHM2018jfm}. This reminds coarsening on soft solid substrates where in an intermediate softness range the translation mode normally dominates \cite{HeST2021sm}. Fig.~\ref{fig:sim_coars_1d_ensemble}~(f) gives the corresponding dependence of the total free energy on time. It shows the typical sequence of step-like transitions between plateaus \cite{PBMT2005jcp}. This can be explained by the observation that transient drop states also correspond to steady states that are however linearly unstable. The plateaus correspond to linear phases where the system slowly leaves an unstable steady state (exponentially slow but accelerating) till strong nonlinear behavior controls the final collapse of one droplet finalizing mass transfer or the coalescence of drops finalizing translation. Ultimately, a single large compound drop in drop-beside-drop configuration remains (Fig.~\ref{fig:sim_coars_1d_ensemble}~(d)). It represents either the state of lowest energy or a metastable state that corresponds to a deep local minimum of the energy.

\begin{figure}[htbp]
	\centering
	\includegraphics[width=0.9\textwidth]{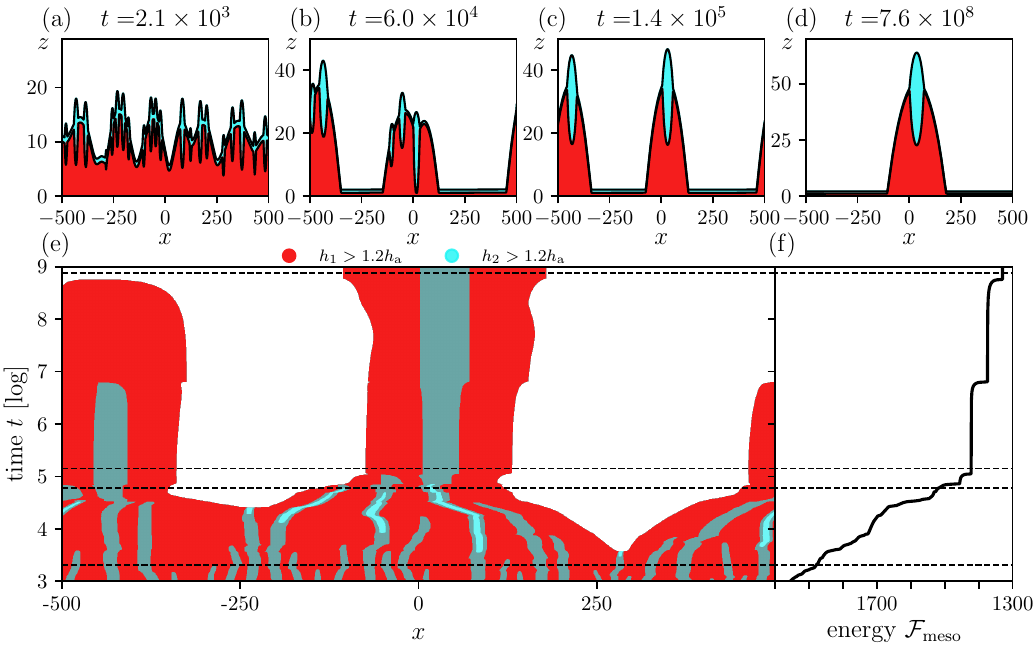}
	\caption{Shown is the coarsening dynamics at  interface energies $(\gam{13},\gam{23},\gamma_\mathrm{s2},\gamma_\mathrm{s3}) = (1.4,1.6,0.8,1.1)$, mean thicknesses $\bar h_{1}=9$ and $\bar h_{2}=3$, and domain size  $L=1000$. Panels, remaining parameters and other details are as in Fig.~\ref{fig:sim_coars_1d_ensemble}. Also, consider the movie (file number 05) in the Supplementary Material \cite{DiTh2024suppmat}.
	\label{fig:sim_coars_1d_ensemble_2}
	}
\end{figure}

The coarsening pathway and emerging final state strongly depend on the system parameters, in particular the energies and mean layer thicknesses. Considering the case of $(\gam{13},\gam{23},\gamma_\mathrm{s2},\gamma_\mathrm{s3}) = (1.4,1.6,0.8,1.1)$ and $\bar{h}_{1}=9$ and $\bar{h}_{2}=3$, we show in Fig.~\ref{fig:sim_coars_1d_ensemble_2} a scenario alternative to Fig. ~\ref{fig:sim_coars_1d_ensemble}. The thinner upper layer results in a fast dewetting of the upper layer on the liquid substrate formed by the lower layer that is however deformed by the process, see Fig.~\ref{fig:sim_coars_1d_ensemble_2}~(a). The resulting many liquid lenses coarsen from nearly 30 to about 10 till dewetting of the lower layer is in full process at $\log t\approx 4.5$, see Fig.~\ref{fig:sim_coars_1d_ensemble_2}~(e). At $t\approx 10^5$ two compound drops have formed, both in nearly symmetric drop-on-drop configuration, see Fig.~\ref{fig:sim_coars_1d_ensemble_2}~(c). As strong asymmetry that promotes translation is missing, they coarsen via mass transfer. Interestingly this occurs in two steps: first, the left compound drop transfers its droplet of liquid~2 to the right compound drop. Second, the remaining droplet of liquid~1 slowly shrinks by mass transfer before it collapses at $t\approx 10^9$. In contrast to Fig. ~\ref{fig:sim_coars_1d_ensemble}, here, the final state is a symmetric drop-on-drop configuration, see Fig.~\ref{fig:sim_coars_1d_ensemble_2}~(d).

\begin{figure}[tbh]
	\includegraphics[width=0.9\textwidth]{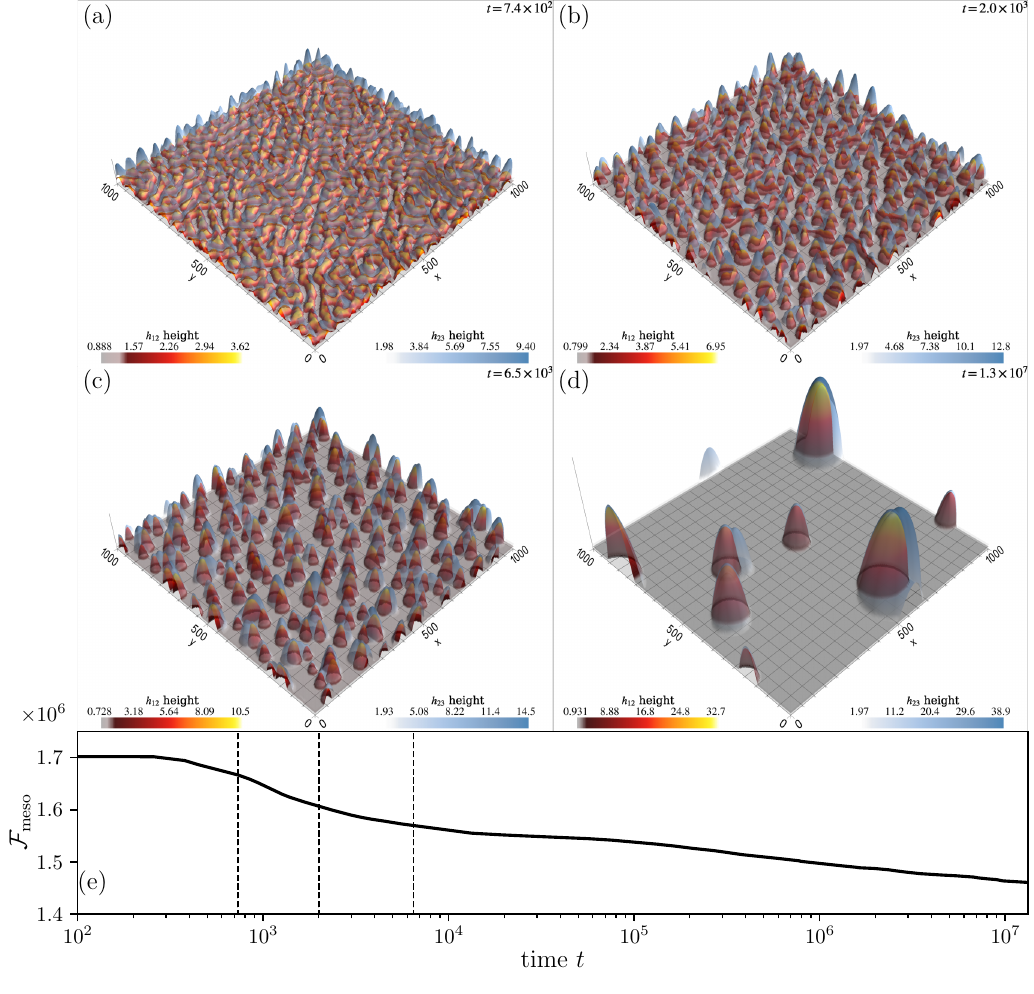}
	\caption{ Shown is the coarsening dynamics for a large ensemble of compound drops on a two-dimensional substrate. Panels~(a) to (d) present selected snapshots at times $t$ marked by dashed vertical lines in the log-normal dependence of total energy on time given in panel (e). Mean thicknesses are $\bar h_{1}=2$ and $\bar h_{2}=2$, and the domain size  is $L\times L$ with $L=1000$. The remaining parameters are as in Fig.~\ref{fig:sim_coars_1d_1drop}. 
	For the visualization of the snapshots we use \textsc{pyvista} \cite{SuKa2019jooss}. 
	Also, consider the Supplementary Material \cite{DiTh2024suppmat} for more snapshots (file number 06) and movies (file numbers 07-11). 
	}
	\label{fig:sim_coars_2d_ensemble}
      \end{figure}
      
Finally, we illustrate  in Fig.~\ref{fig:sim_coars_2d_ensemble} the usage of the model for studying the coarsening of compound drops on two-dimensional substrates, thereby  using the same parameters as in Fig.~\ref{fig:sim_coars_1d_ensemble}. After a short dewetting phase with a labyrinthine stripe-like structure (Fig.~\ref{fig:sim_coars_2d_ensemble}~(a), also cf.~Fig.~8 of Ref.~\cite{PBMT2006el}), compound droplets emerge and start to coarsen (Fig.~\ref{fig:sim_coars_2d_ensemble}~(b)). Due to the large number of droplets, individual coarsening events can not be discerned in the smooth decrease of the total energy with advancing time, see Fig.~\ref{fig:sim_coars_2d_ensemble}~(e). Careful inspection of the intermediate panel Fig.~\ref{fig:sim_coars_2d_ensemble}~(b) already shows the prevalence of asymmetric drop-beside-drop configurations. However, also other (transient) configurations can be spotted: compound drops with more than two subdrops (two or three drops of liquid~1 covered by or adjacent to a larger drop of liquid~2), drop-on-ring structures mostly with partial rings unlike more symmetric transient states shown in Fig.~10 of \cite{DGGR2021l}. These complex configurations have mostly disappeared in  Fig.~\ref{fig:sim_coars_2d_ensemble}~(c) where drop-beside-drop configurations dominate. When only a few drops are left they mostly correspond to the ultimately surviving drop-beside-drop configuration, see Fig.~\ref{fig:sim_coars_2d_ensemble}~(d). This agrees with the case of the one-dimensional substrate in Fig.~\ref{fig:sim_coars_1d_ensemble}. Note that due to the large numerical effort caused by exponentially increasing coarsening times we do not follow the process till the state of minimal energy is reached.
\section{Conclusion}
\label{sec:conc}
We have revisited the mesoscopic hydrodynamic description of the dynamics of sessile compound drops consisting of two immiscible nonvolatile partially wetting liquids on horizontal and inclined homogeneous smooth rigid solid substrates. We have discussed that there exist two formulations (in terms of either {layer thickness or interface height} profiles) that transparently correspond to a (thermodynamic) gradient dynamics form. {A third one that mixes the two previous formulations should be avoided as it obfuscates the gradient dynamics form and complicates a linear stability analysis because} the resulting mobility matrix is asymmetric. The subsequent discussion of the wetting energy has clarified that in contrast to common praxis three {contributions} are needed (interaction across lower layer, interaction across upper layer, and interaction across both layers - resulting in three Hamaker constants) to be able to reflect the full spectrum of macroscopic parameters {for partially wetting liquids} in the mesoscopic model. A condition for general wetting energies, $g(h_1,h_2)\neq g_1(h_1)+g_2(h_2)$, has also been given. 
Based on these considerations, we have determined corresponding consistency relations between macroscopic and mesoscopic parameters that ensure the full equivalence of all mesoscopic Laplace, Neumann and Young laws with their macroscopic counterparts. They are most conveniently established by employing the full-curvature variant of the model but fully carry over to the long-wave variant. In this way, we have corrected a number of oversights in {some} previous models that were intended to capture the full parameter spectrum {for partially wetting liquids} but did not. {Here we remark that our focus on partially wetting liquids mesoscopically described by a wetting energy that consists of three terms of identical functional form allows for transparent consistency conditions between mesoscopic and macroscopic description. These considerations have to be amended if a wider class of wetting energies is considered. This includes, e.g., situations where one of the liquids is wetting or situations where first order wetting transitions can occur.}

The resulting dynamic model now allows one to consider dynamic processes involving two-layer films and compound drops consisting of two immiscible {partially wetting} liquids on solid substrates. Our approach ensures that these processes ultimately converge to equilibrium states that can be exactly predicted based on macroscopic parameters. It can also be employed to investigate how the behavior is influenced by mesoscale effects when considering very small drops. Here, we have considered three examples: the spreading of individual compound drops on a one-dimensional horizontal substrate, sliding compound drops on one-dimensional inclined substrates (by incorporating the 'downhill-component' of gravity into the model), and the coarsening of drop ensembles on one- and two-dimensional horizontal substrates.

First, we have investigated the spreading of compound drops employing a symmetric initial configuration where a drop of liquid~2 completely covers a drop of liquid~1 (drop-covers-drop configuration). We have found that even at fixed interface energies, the dynamic behavior may dramatically depend on the volume ratio of the two liquids and the initial contact angles: The outer drop may spread faster than the inner one resulting in no change of configuration. However, if the inner drop spreads faster the compound drop can change configuration. When the inner contact line reaches the vicinity of the outer one, the latter can either slightly recede before a Neumann region develops and the configuration becomes one where a drop of liquid 2 only covers the central top part of the drop of liquid 1 (drop-on-drop configuration), or the contact lines meet forming a four-phase contact. This four-phase configuration can represent the final stable state or only form a short or long transient intermediate configuration before a drop-on-drop configuration emerges. The discussed transitions all occur between configurations discussed as possible equilibrium states in Refs.~\cite{MaAP2002jfm,PBMT2005jcp,NTGD2012sm,BSNB2014l,ZCAG2016jcis,Kita2024jons}. Note, in particular, {Ref.}~\cite{MaAP2002jfm} for the four-phase configuration that our simulations consistently confirm as the final equilibrium state in a finite range of control parameters. This seems to indicate that morphological phase diagrams as in Ref.~\cite{NTGD2012sm} might need to be reconsidered in the relevant parameter ranges. Note that the gradient dynamics formulation has allowed us to study the importance of the different dissipation channels and corresponding changes at morphological transitions.

Second, we have considered drops sliding on a homogeneous incline. Also here, transitions to other configurations may occur: Obviously an initially symmetric drop-covers-drop configuration will become asymmetric due to the down-hill driving force that breaks the left-right symmetry of the system. However, we have not encountered an asymmetric drop-covers-drop configuration as stationary state, i.e., a compound drop that slides with constant speed and shape. Instead, the final configuration might be a drop of liquid 1 adjacent in front of a drop of liquid 2 or vice versa (drop-beside-drop configuration). Also an initial drop-on-drop configuration will transition into a drop-beside-drop configuration. Here, a more extensive parametric study will most likely uncover more possible transitions. Note that also here the gradient dynamics formulation allows one to study the influence of the different dissipation channels and their dependence on parameters. We have left this as a task for the future.

Finally, we have returned to horizontal substrates and studied the coarsening of droplet ensembles on one- and two-dimensional substrates. We find that the two basic coarsening models of mass transfer between drops and translation of drops both contribute. Their relative importance depends not only on drop size but also on configuration, an effect that requires further future scrutiny. Note that also in the two-dimensional case, asymmetric drops do not only appear as long-lived transients but may also feature as final equilibrium configuration - in agreement with \cite{NTGD2012sm,BSNB2014l}, termed \enquote{Janus drops} in \cite{NTGD2012sm}. Note that the existence of such states is negated in Ref.~\cite{MaAP2002jfm} based on a torque balance argument that, however, does not apply in situations without an external force field like gravity. In \cite{PBMT2005jcp,PBMT2006el} all drop-like long-lived transients and final states are symmetric {while in \cite{Kita2024jons} asymmetric intermediate states are found}. Also here future studies should expand on \cite{Kita2024jons,Diekmann2022Munster} and provide morphological phase diagrams that clarify the ranges of existence and stability of such states {in terms of physical control parameters. We are not aware of recent experimental work on large-scale dewetting and subsequent coarsening phenomena for bi- or multi-liquid systems of partially wetting liquids beyond the already older works on decomposing and dewetting films reviewed in \cite{GeKr2003pps,CrMa2009rmp}. We hope our study motivates further experiments on dewetting phenomena with a focus on complex liquids and multi-liquid set-ups. An interesting avenue could be the pulsed-laser-induced dewetting of multilayer metal films similar to the single-layer set-ups reviewed in \cite{KGDF2020arfm}. Also see \cite{DGGR2021l} for decomposing dewetting nanoscopic films of alloys.}

Furthermore, our gradient dynamics approach lends itself to a number of future amendments and extensions. Beside the relatively simple incorporation of topographically and chemically heterogeneous substrates like for simple liquids done in \cite{TBBB2003epje,SaKa2012jem}, e.g., to investigate the stick-slip motion \cite{ThKn2006prl} of compound drops, one may in a straightforward manner incorporate hydrostatic effects including different mass densities for the liquids.
Beside gravity or electrical fields \cite{LKRS2002m,ABCO2012sm,BaSh2007jcis} one might incorporate volatility \cite{HDJT2023jfm}, soft elastic substrates \cite{HEHZ2022prsa}, or surfactants \cite{ThAP2016prf}. Although the systems then become rather complicated, employing the gradient dynamics approach to mesoscopic hydrodynamics \cite{Thie2018csa} allows for the combination of building blocks known from simpler systems. However, one has to go beyond the gradient dynamics framework if thermocapillary effects shall be added, e.g.,  to study the influence of internal convection and long-wave {thermal} Marangoni instabilities \cite{OrRo1992jpif,ThKn2004pd}. For two-layer films such influences were incorporated into the two-layer model (but not employed) in \cite{PBMT2005jcp}, for further studies see \cite{NeSi2007pf,NeSi2009prl,NeSi2017pf,NeSi2021prf}. Here, building on the present model would allow one to assess how thermocapillarity influences Young and Neumann angles.

\acknowledgments

The authors acknowledge support by the Deutsche Forschungsgemeinschaft (DFG) via Grant No.\ TH781/12-1 and TH781/12-2 within Priority Program (SPP)~2171 \enquote{Dynamic Wetting of Flexible, Adaptive, and Switchable Surfaces}; {We acknowledge preliminary work of Kevin Mitas on earlier models for two-layer systems published in his PhD thesis. Furthermore, we} acknowledge fruitful discussions with Daniel Greve, Simon Hartmann, Christopher Henkel and Dominik Thy at the University of M\"unster, and with participants of the events organized by SPP~2171. Part of the calculations were performed on the HPC cluster PALMA II of the University of Münster, subsidized by the DFG via Grant INST 211/667-1.
\newline

This version of the article has been accepted for publication, after peer review but
is not the Version of Record and does not reflect all post-acceptance improvements and
corrections. The Version of Record is available online at \href{https://doi.org/10.1103/PhysRevFluids.10.024002}{10.1103/PhysRevFluids.10.024002}.
\section*{Data availability statement}
The underlying data and source code required to reproduce the shown results is publicly available at the
data repository \textit{zenodo} with the corresponding \textsc{doi}:\href{https://doi.org/10.5281/zenodo.13143462}{10.5281/zenodo.13143462}.
\appendix

\section{Macroscopic description}
\label{sec:statics-macro}
For convenience of the reader, here, we review the macroscopic description of static compound drops on rigid solid substrates. In particular, we introduce the energy functional $\mathcal{F}_\text{macro}$ based on the six interface energies (liquid~1 to fluid~3 (here ambient gas, but may also be an ambient liquid) $\gamma_{13}$, liquid~2 to fluid~3 $\gamma_{23}$, liquid~1 to liquid~2 $\gamma_{12}$, liquid~1 to substrate $\gamma_{\mathrm{s}1}$, liquid~2 to substrate $\gamma_{\mathrm{s}2}$, and fluid~3  to substrate $\gamma_{\mathrm{s}3}$), and summarize how minimization of the total energy results in Laplace, Neumann, and Young laws and point out relations between them. This forms the basis of establishing the consistency conditions discussed in section~\ref{sec:statics-consistency}.

To discuss the energy, we consider a drop-on-drop configuration consisting of a drop of liquid~2 (of volume $V_2$ and lateral extension $2\ltwo$) on top of a drop of liquid~1 (of volume $V_1$ and lateral extension $2\lone$) on a one-dimensional rigid smooth solid substrate as sketched in Fig.~\ref{fig:sketch_macro_meso}~(a). 
The configuration is symmetric and is embedded in an ambient fluid~3.
The configuration features five of the six possible interfaces, indicated by the respective interface energies $\gamma_{ij}$, e.g., $\gam{12}$ at the interface 1-2. 
The three static thickness profiles $\hontw,\,\honth$, and $\htwth$ correspond to circular arcs. 
An alternative drop-covers-drop configuration (with a much larger $V_2$) would also feature the sixth energy $\gam{s2}$ (cf.~Fig.~\ref{fig:sketch_macro_meso}(c)).
\subsection{Energy functional}
The macroscopic energy functional for a compound drop consisting of two immiscible liquids for the configuration in Fig.~\ref{fig:sketch_macro_meso}(a) is given by
\begin{align}
	\begin{split}
		\mathcal{F}_\text{macro}= & 
		\int_{0}^{\lone} \gsl \mathrm{d}x +
		\int_{0}^{\ltwo} (\gontw \xiontw + \gtwth \xitwth) \mathrm{d}x +
		\int_{\ltwo}^{\lone} \gonth \xionth \mathrm{d}x  +
		\int_{\lone}^{D} \gsg \mathrm{d}x \\
		&-P_1\left( \int_{0}^{\ltwo} \hontw \mathrm{d}x  + \int_{\ltwo}^{\lone}\honth\mathrm{d}x -V_1\right) 
		-P_2\left(\int_{0}^{\ltwo}\left( \htwth - \hontw \right) \mathrm{d}x -V_2\right)\\
		&+ \lambda_{1} \honth\left(\lone\right)
		+\lambda_{2} \left( \htwth\left(\ltwo\right) - \hontw\left(\ltwo\right)\right)
		+ \lambda_{3}\left( \htwth\left(\ltwo\right) - \honth\left(\ltwo\right)\right) \label{eq:macro_energy_lagrangemult}\,,
	\end{split}
\end{align}
where due to the reflection symmetry we only consider half the domain, i.e., $x\in[0,D]$, assume that all interface profiles can be written in Monge representation $h_{ij}(x)$, and use the one-dimensional equivalents of the metric factors in the mesoscopic energy~\eqref{eq:meso_grandpotential}, i.e., $\xi_{ij}=\sqrt{1 + \left(\frac{\mathrm{d}h_{ij}}{\mathrm{d}x}\right)^2}$. The constants $P_1,$ $P_2$, $\lambda_{1}$, $\lambda_{2}$ and $\lambda_{3}$ represent the Lagrange multipliers for various constraints, namely, controlling the volumes of liquids~1 and~2
\begin{align}
	\int_{0}^{\ltwo} \hontw \mathrm{d}x + \int_{\ltwo}^{\lone} \honth \mathrm{d}x  &= V_1,\label{eq:macro_masscon1}\\
	\int_{0}^{\ltwo} (\htwth-\hontw)\mathrm{d}x &= V_2, \label{eq:macro_masscon2}
\end{align}
and ensuring geometric consistency
\begin{align}
	\honth\left(\lone\right)&=0 \label{eq:macro_geomconsis1},\\
	\hontw\left(\ltwo\right)&=\htwth\left(\ltwo\right) \label{eq:macro_geomconsis2},\\
	\honth\left(\ltwo\right) &= \htwth\left(\ltwo\right)\label{eq:macro_geomconsis3},
\end{align}
respectively.
Next, the energy \eqref{eq:macro_energy_lagrangemult} is minimized w.r.t.\ variations in the profiles $h_{ij}(x)$ and the positions of the contact lines $L_1$ and $L_2$.

\subsection{Minimization of $\mathcal{F}_\text{macro}$}
Minimization of the energy functional \eqref{eq:macro_energy_lagrangemult} w.r.t.\ the profiles $h_{12},\,h_{23}$ and $h_{13}$ gives
\begin{align}\label{eq:macroI_minim_h12}
	\delta \mathcal{F}_\text{macro} = 
	\left(\gontw \dfrac{\partial_x \hontw}{\xiontw} - \lambda_2\right)\delta \hontw\left(\ltwo\right)
	- \int_{0}^{\ltwo}\left(\gontw\dfrac{\partial_{xx}\hontw}{\xiontw^3} + (P_1-P_2)\right) \delta \hontw(x)\mathrm{d}x,
\end{align}
\begin{align}\label{eq:macroI_minim_h23}
	\delta \mathcal{F}_\text{macro} = 
	\left(\gtwth \dfrac{\partial_x \htwth}{\xitwth} + \lambda_2 + \lambda_3\right)\delta \htwth\left(\ltwo\right)
	- \int_{0}^{\ltwo}\left(\gtwth\dfrac{\partial_{xx}\htwth}{\xitwth^3} + P_2\right) \delta \htwth(x)\mathrm{d}x,
\end{align}
and
\begin{align}\label{eq:macroI_minim_h13}
	\begin{split}
		\delta \mathcal{F}_\text{macro} =&
		\left(\gonth \dfrac{\partial_x\honth}{\xionth} + \lambda_1 \right) \delta \honth\left(\lone\right)
		- \left(\gonth \dfrac{\partial_x\honth}{\xionth} + \lambda_3\right) \delta \honth\left(\ltwo\right)\\
		\,&-\int_{\ltwo}^{\lone} \left(\gonth \dfrac{\partial_{xx}\honth}{\xionth^3} + P_1\right) \delta \honth(x)\mathrm{d}x,
	\end{split}
\end{align}
respectively,
while minimization w.r.t.\ the contact line positions results in 
\begin{align}\label{eq:macroI_minim_l1}
	\delta \mathcal{F}_\text{macro} = \left(\gonth \xionth + \gsl - \gsg - P_1 \honth + \lambda_1 \partial_x\honth\right)\delta\lone \quad\mathrm{at}\quad x=L_1
\end{align}
and
\begin{align}\label{eq:macroI_minim_l2}
	\begin{split}
		\delta \mathcal{F}_\text{macro} =&\left(
		\gontw\xiontw
		+\gtwth\xitwth
		-\gonth\xionth
		-P_2\left(\htwth-\hontw\right)
		-P_1\left(\hontw-\honth\right)\right.\\
		\,& \left.
		+\lambda_2 \left(\partial_x\htwth - \partial_x\hontw\right)
		+\lambda_3\left(\partial_x\htwth-\partial_x\honth\right)
		\right)\delta \ltwo \quad\mathrm{at}\quad x=L_2.
	\end{split}
\end{align}

We identify the expression $\frac{\partial_{xx} h_{ij}}{\xi_{ij}^3}=\kappa_{ij}$ as the local curvature of $h_{ij}(x)$. 

\subsection{Laplace, Neumann and Young laws}\label{app:neumann}

We then obtain from Eqs.~\eqref{eq:macroI_minim_h12} to~\eqref{eq:macroI_minim_h13} the Laplace laws
\begin{align}
	\gontw \kappa_{12}&= P_2-P_1 &&\text{for } x\in [0, \ltwo] \label{eq:macro_laplace12}\\
	\gtwth \kappa_{23} &= -P_2 &&\text{for } x\in [0, \ltwo] \label{eq:macro_laplace23}\\
	 \gonth \kappa_{13}&= - P_1 &&\text{for } x\in [\ltwo, \lone] \label{eq:macro_laplace13}
\end{align}
and the relations for the Lagrange multipliers
\begin{align}
	\gontw \dfrac{\partial_x\hontw}{\xiontw} &= \lambda_{2}  &&\text{at } x=\ltwo \label{eq:macroI_neumannI}\,,\\
	\gtwth \dfrac{\partial_x\htwth}{\xitwth} &= -\lambda_{2} - \lambda_{3}  &&\text{at } x=\ltwo \label{eq:macroI_neumannII}\,,\\
	\gonth\dfrac{\partial_x\honth}{\xionth} &= -\lambda_{3}  &&\text{at } x=\ltwo \label{eq:macroI_neumannIII} \\
	\gonth \dfrac{\partial_x\honth}{\xionth} &= -\lambda_{1}  &&\text{at } x=\lone \label{eq:macroI_neumannIIb}\,.
\end{align}
The Laplace laws state that the interfaces have constant curvatures $\kappa_{ij}$, i.e., circular arcs in the considered geometry. Thereby, the $\kappa_{ij}$ are determined by the pressure jumps across the corresponding interface (identifying $P_1$ and $P_2$ as the pressure in liquid ~1 and~2, respectively). Care has to be taken as different sign conventions exist. Combining Eqs.~\eqref{eq:macro_laplace12} to \eqref{eq:macro_laplace13} gives the relation
\begin{align}
  0 = -\gonth \kappa_{13} + \gontw \kappa_{12} + \gtwth \kappa_{23}.
  \label{eq:relation-cirvatures}
\end{align}

Next, we consider the three-phase contact of the three fluids 1, 2, and 3 as shown in Fig.~\ref{fig:sketch_macro_meso}~(a), i.e., Neumann angles are defined as angles between the tangent to the respective interface at the contact point and the horizontal. 
We relate the angles to the interface slopes by $\partial_x h_{ij}\left.\right|_{x=\ltwo}=\pm \tan(\neum{ij})$ the ``+'' is for interface 1-2, and the  ``-'' is for interfaces 2-3 and 1-3, i.e., all angles shown in Fig.~\ref{fig:sketch_macro_meso}~(a) are defined positive. 

The minimizations give Young's law for the three-phase contact of substrate, liquid~1, and fluid~3 at $x=L_1$  and Neumann's law at the three-fluid contact point at  $x=L_2$. 
Combining  (\ref{eq:macroI_neumannI}), (\ref{eq:macroI_neumannII}) and (\ref{eq:macroI_neumannIII}) gives
\begin{align}
	\gtwth \dfrac{\partial_x\htwth}{\xitwth} &= -\gontw \dfrac{\partial_x\hontw}{\xiontw} + \gonth \dfrac{\partial_x\honth}{\xionth}\,.
\end{align}
Employing $\xi_{ij}^{-1}\left.\right|_{x=\ltwo}=\cos(\neum{ij})$, this becomes the vertical component of Neumann's law
\begin{align}
  \gtwth \sin(\neum{23}) &=\gontw \sin(\neum{12}) + \gonth \sin(\neum{13})
                           \label{eq:macroI_neumannvert}.
\end{align}
Further, we use Eq.~\eqref{eq:macroI_minim_l2} in the limit $P_1, P_2\to 0$ (i.e., for large drops) together with Eqs.~\eqref{eq:macroI_neumannI} and \eqref{eq:macroI_neumannIII} to get 
\begin{align}
	\begin{split}
		0=&\frac{\gontw }{\cos(\neum{12})} +  \frac{\gtwth}{\cos(\neum{23})} -  \frac{\gonth}{\cos(\neum{13})} \\
		&+\gontw \sin(\neum{12}) (-\tan(\neum{23})- \tan(\neum{12})) + \gonth\sin(\neum{13}) (-\tan(\neum{23})+\tan(\neum{13}))\,.
	\end{split}
\end{align}
Using Eq.~\eqref{eq:macroI_neumannvert} and rearranging the trigonometric functions we obtain the horizontal component of Neumann's law
\begin{align}
\gonth \cos(\neum{13})=& \gontw \cos(\neum{12}) +\gtwth \cos(\neum{23}). \label{eq:macroI_neumannhoriz}
\end{align}
Finally, Young's law is obtained rearranging Eqs.~\eqref{eq:macroI_minim_l1} and \eqref{eq:macroI_neumannIIb} into 
\begin{align}
\gonth \cos(\young{13}) &= \gsg - \gsl\,. \label{eq:macroI_younglaw13}
\end{align}
Considering other compound drop configurations than the one in Fig.~\ref{fig:sketch_macro_meso}(a), e.g., the one in Fig.~\ref{fig:sketch_macro_meso}(c), one obtains similar Young laws for contact points of the substrate with on the one hand liquid~2 and fluid~3 {(cf.~Fig.~\ref{fig:sketch_macro_meso}(c), outer contact line)},
\begin{align}
\gamma_{23} \cos(\young{23}) &= \gamma_{\mathrm{s}3} - \gamma_{\mathrm{s}2}\,, \label{eq:macroI_younglaw23}
\end{align}
and on the other hand with liquid~1 and liquid~2 {(cf.~Fig.~\ref{fig:sketch_macro_meso}(c), inner contact line)},
\begin{align}
\gamma_{12} \cos(\young{12}) &= \gamma_{\mathrm{s}2} - \gamma_{\mathrm{s}1}\,. \label{eq:macroI_younglaw12}
\end{align}
Note that Eqs.~\eqref{eq:macroI_younglaw13}-\eqref{eq:macroI_younglaw12} can be combined into 
\begin{align}
	0= -\gonth \cos(\theta_{13}) + \gontw\cos(\theta_{12}) + \gtwth\cos(\theta_{23})\,,
\end{align}
a relation somewhat similar to \eqref{eq:relation-cirvatures} and \eqref{eq:macroI_neumannhoriz}. 

\subsection{Long-wave approximation}
\label{sec:long-wave-macro}

Finally, we consider the long-wave approximation of the macroscopic relations obtained in section~\ref{app:neumann}.
To do so, we assume that all Young and Neumann angles are small, i.e., we approximate $\cos\phi\approx 1-\phi^2/2$ and $\sin\phi\approx\phi$. In this way, the components \eqref{eq:macroI_neumannvert} and \eqref{eq:macroI_neumannhoriz} of Neumann's law  become
\begin{align}
	 \gontw \neum{12} + \gonth \neum{13} &= \gtwth \neum{23},		 \label{eq:macroI_neumannvert-lw}\\
\text{and }\qquad\gontw \neum{12}^2 + \gtwth \neum{23}^2 -\gonth \neum{13}^2&= 2(\gontw +\gtwth -\gonth)\,, \label{eq:macroI_neumannhoriz-lw}
\end{align}
respectively.
The Young laws \eqref{eq:macroI_younglaw13}, \eqref{eq:macroI_younglaw23} and \eqref{eq:macroI_younglaw12} become
\begin{align}
\gonth \young{13}^2 &= 2(\gsl + \gonth - \gsg) \,, \label{eq:macroI_younglaw13-lw}\\
\gamma_{23} \young{23}^2 &= 2(\gamma_{\mathrm{s}2} + \gamma_{23} - \gamma_{\mathrm{s}3})  \,, \label{eq:macroI_younglaw23-lw}\\
\text{and }\qquad\gamma_{12} \young{12}^2 &= 2(\gamma_{\mathrm{s}1}+\gamma_{12} -\gamma_{\mathrm{s}2}) \,, \label{eq:macroI_younglaw12-lw}
\end{align}
respectively.
By combining them one obtains the relation
\begin{align}
\gamma_{12} \young{12}^2+\gamma_{23} \young{23}^2 -\gonth \young{13}^2 = 2(\gamma_{12}+\gamma_{23}-\gamma_{13})
\end{align}
again a structure very similar to the horizontal component of Neumann's law \eqref{eq:macroI_neumannhoriz-lw}.

\section{Related physical quantities for the mesoscopic model}
\label{sec:app:supp}
\subsection{Velocitiy fields}
\label{app:velocity_fields} 
To reconstruct the vertically resolved velocity fields $\vec{v}_1=(v_1, w_1)^T$ and $\vec{v}_2=(v_2, w_2)^T$ of the lower and upper liquid layer, respectively, from the film height profiles one may perform the standard derivation of the thin-film equation for the two-layer system \cite{OrDB1997rmp,PBMT2005jcp,CrMa2009rmp,JPMW2014jem}. 
Explicitly, one uses no-slip and no-penetration boundary conditions at the solid substrate, continuity of shear stress and velocity at the liquid 1-liquid 2 interface, and zero shear stress and the kinematic boundary condition at the liquid 2-gas interface. Using the formulation in terms of layer thicknesses, one ultimately obtains the form in Eq.~\eqref{eq:mobilitymatrix}. However, as an intermediate result, one directly obtains the height-resolved $x$-components of the velocities $v_i(x,z,t)$ in the two liquid layers. The  $z$-components one obtains using the incompressibility condition $\partial_x v_i + \partial_z w_i=0$ and integration in $z$ (similar to~Ref.~\cite{EWGT2016prf}) to finally obtain
\begin{align}
	v_{1} =& 
		\frac{1}{\eta_1}\left[\frac{z^2}{2}\partial_x p_1- z (h_1 \partial_x p_1 +  h_2  \partial_x p_2)\right]
		\\
	w_{1}=& 
		-\frac{1}{2\eta_1}\left[ \frac{z^3}{3} \partial_{xx}p_1 - z^2 \partial_{x}(h_1 \partial_x p_1 + h_2 \partial_x p_2)\right] 
		\,,
\end{align}
and 
\begin{align}
	v_{2} =& \frac{1}{\eta_2}\left[\frac{z^2}{2}\partial_x p_2 - z (h_1+h_2)\partial_x p_2 + I\right] 
		\\
	\begin{split}
	w_{2}=& -\frac{1}{\eta_2}\left[ 
            \frac{z^3}{6}\partial_{xx}p_2 - \frac{z^2}{2}\partial_x\left[(h_1+h_2)\partial_xp_2 \right]+ z \partial_x I \right.\\
    &\qquad\quad\left. -\frac{h_1^3}{6}\partial_{xx}p_2 + \frac{h_1^2}{2}\partial_x\left[(h_1+h_2)\partial_xp_2 \right]- h_1 \partial_x I - \eta_2 w_1|_{z=h_1}
    \right]\,.
	\end{split}
\end{align}
Here, $I=\frac{h_1^2}{2}\partial_x (p_2 - p_1 \eta_2/\eta_1)+(1-\eta_2/\eta_1)h_1h_2\partial_xp_2$, and the pressures $p_1$ and $p_2$ correspond to the ones in Eq.~\eqref{eq:BCP-ENRG}. 

\subsection{Dissipation channels}
\label{app:dissipation}
In order to analyze how dissipation occurs via the different channels, we consider the total time derivative of the free energy in Eq.~\eqref{eq:meso_grandpotential}, i.e., the negative of the total dissipation $D$ and split it into parts $D_1$ and $D_2$ corresponding to liquid~1 and liqud~2, respectively. The approach is similar to the one employed in \cite{GrHT2023sm, HDGT2024l} for drops on adaptive substrates. Using the formulation in terms of layer thicknesses in Eqs.~\eqref{eq:pde_system}-\eqref{eq:mobilitymatrix} allows us to directly determine $D_1$ and $D_2$.  For the one-dimensional domain we obtain
\begin{align}
\frac{\mathrm{d}}{\mathrm{d}t}\mathcal{F} &= \int_\Omega \sum_{\alpha=1}^2 \frac{\delta \mathcal{F}}{\delta  h_\alpha}\partial_t h_\alpha \mathrm{d}x
= \int_\Omega \sum_{\alpha=1}^2 \frac{\delta \mathcal{F}}{\delta  h_\alpha}(-\partial_xj_\alpha) \mathrm{d}x
= \sum_{\alpha=1}^2 \int_\Omega  \left(\partial_x\frac{\delta \mathcal{F}}{\delta  h_\alpha}\right) j_\alpha\mathrm{d}x
\label{eq:diss_1}
\\
&= -\int_\Omega \left(\partial_x \frac{\delta \mathcal{F}}{\delta  h_1}\right) \left[\sum_{\beta=1}^2  Q_{1\beta}\partial_x\frac{\delta \mathcal{F}}{\delta  h_\beta}\right]
\mathrm{d}x
 -\int_\Omega
 \left(\partial_x \frac{\delta \mathcal{F}}{\delta  h_2}\right) \left[\sum_{\beta=1}^2  Q_{2\beta}\partial_x\frac{\delta \mathcal{F}}{\delta  h_\beta}\right]
\mathrm{d}x \,.\label{eq:diss_2}
\end{align}
In the final transformation of Eq.~\eqref{eq:diss_1} we have used partial integration and eliminated the surface terms by assuming periodic boundary conditions. The negative of the first and second term on the right hand side of \eqref{eq:diss_2} represent the two dissipation channels, namely, the overall viscous dissipation $D_1$ and $D_2$ in liquid~1 and in liquid~2, respectively.

\end{document}